\documentclass[lettersize,journal]{IEEEtran} 

\usepackage{amsmath,amsfonts}
\usepackage{algorithmic}
\usepackage{algorithm, color}
\usepackage{array}
\usepackage[caption=false,font=normalsize,labelfont=sf,textfont=sf]{subfig}
\usepackage{textcomp}
\usepackage{stfloats}
\usepackage{url}
\usepackage{verbatim}
\usepackage{graphicx}
\usepackage{cite}

\usepackage{amssymb,amsmath,epsfig,graphicx,theorem} 
\usepackage{arydshln}
\usepackage{cite}
\newtheorem{Theo}{Theorem}

\newtheorem{lemma}{Lemma}
\newtheorem{remark}{Remark}  
\allowdisplaybreaks 

\begin{document}

\title{Capacity of Hierarchical  Secure Coded Gradient Aggregation with  Straggling Communication Links}

\author{Qinyi Lu, Jiale Cheng, Wei Kang and Nan Liu 
        
\thanks{
	Q. Lu and W. Kang are with the School of Information Science and Engineering, Southeast University, Nanjing 211189, China. 
	(e-mail: \{qylu, wkang\}@seu.edu.cn).  
	J. Cheng is with EECS Department of the University of Michigan Ann Arbor, MI 48109, USA.
	(e-mail: jlcheng@umich.edu).  
	N. Liu is with the National Mobile Communications Research Laboratory, Southeast University, Nanjing 211189, China. (e-mail: nanliu@seu.edu.cn). }}



\maketitle

\begin{abstract} 
The growing privacy concerns in distributed learning have led to the widespread adoption of secure aggregation techniques in distributed machine learning systems, such as federated learning.    
Motivated by a coded gradient aggregation problem in a user-helper-master hierarchical network setting with straggling communication links, we formulate a new secure hierarchical coded gradient aggregation problem. 
In our setting, \( K \) users communicate with the master through an intermediate layer of \( N \) helpers, who can communicate with each other.   
With a resiliency threshold of \( N_r \) for straggling communication links, and at most \( T \) colluding helpers and any number of colluding users, the master aims to recover the summation of all users' gradients while remaining unaware of any individual gradient that exceeds the expected summation.  
In addition, helpers cannot infer more about users' gradients than what is already known by the colluding users.
We propose an achievable scheme where users' upload messages are based on a globally known Vandermonde matrix, and helper communication is facilitated using an extended Vandermonde matrix with special structural properties. 
A matching converse bound is also derived, establishing the optimal result for this hierarchical coded gradient aggregation problem.  
\end{abstract} 


\section{Introduction}
With the widespread use of mobile devices and the continuous growth of data, distributed learning, particularly federated learning \cite{McMahan2016, Li2020}, has emerged as a promising approach. 
In distributed learning systems, a large number of mobile users collaboratively train a global model using locally collected datasets, with the local updates  being aggregated by a central server.  
 
As revealed in \cite{zhu2019deep}, the uploading of users' local updates can expose sensitive information.   
Aiming to enhance security and enable collaborative model training, Bonawitz \emph{et al.} introduced a practical secure aggregation protocol, referred to as \emph{SecAgg}, in \cite{2017Practical}.   
In the SecAgg protocol, pairwise random masks between users ensure the security of individual users and enable the central server to obtain the desired aggregation, with the pairwise random masks from different users  canceling each other out.  
A series of subsequent studies have focused on developing improved secure aggregation protocols, aiming to achieve higher communication efficiency, accommodate diverse security constraints, enhance robustness against user straggling, and address other related challenges \cite{so2022lightsecagg,2021SoJTurbo,2022SunHua,2022JahaniSwiftAgg, jahani2023swiftagg+,Zhao2024,Wan2024,Wan20242,so2023securing,2021SoJBuffered,2022Nguyen,2020KadheFastSecAgg,2020Beongjun, 2022ElkordyHeteroSAg, 2020James,UsSami2024}.   
More specifically, So, \emph{et al.} focused on robustness against user straggling by employing efficient methods, such as random mask reconstruction for dropped users or turbo aggregation, thereby reducing the communication overhead introduced by the masks \cite{so2022lightsecagg,2021SoJTurbo}.  
Under the information-theoretic system model framework, the secure aggregation problem with straggling and colluding users \cite{2022SunHua,2022JahaniSwiftAgg, jahani2023swiftagg+}, as well as groupwise keys \cite{Zhao2024,Wan2024,Wan20242}, has been studied.
When the focus is not only on a single round of training in distributed learning, problems including multi-round privacy leakage \cite{so2023securing} and buffered asynchronous aggregation \cite{2021SoJBuffered,2022Nguyen} have been extensively studied.
 
The secure aggregation problems mentioned above are primarily based on a simple user-server network architecture.
However, in practical applications, mobile users may be located at a considerable distance from the central server, resulting in high latency and low bandwidth.  
As a result, hierarchical distributed learning systems have garnered widespread research attention \cite{liu2020,luo2024,wang2022,Chen2022,tang10_2024,Prakash2020,Liang2020,Sasidharan2021,Krishnan2023}. 
More specifically,  Prakash \emph{et al.} investigate the coded gradient aggregation problem in a hierarchical setup, where multiple users collaborate with reliable helper nodes to assist in gradient aggregation at the master node, as described in \cite{Prakash2020}.  This study, along with subsequent works \cite{Liang2020, Sasidharan2021, Krishnan2023}, focuses on the robustness of the straggling user-to-helper links and explores methods to minimize the communication load between users and helpers, as well as between helpers and the master node.

In this paper, we investigate a hierarchical coded gradient aggregation (HCGA) problem similar to those in \cite{Prakash2020, Sasidharan2021, Krishnan2023}, while also incorporating additional security constraints to protect user data privacy.  
Secure aggregation in hierarchical distributed learning systems has been explored in \cite{Egger2023,eggerarxiv2024,zhang2024optimalcommunicationkeyrate,tavallaie2024}.  
Unlike the existing works on hierarchical secure aggregation mentioned above, we consider a different hierarchical network model, where each user is connected to multiple helpers and focus on the robustness of the straggling user-to-helper links and helper-to-master links. 
Such different assumptions can be found in practical applications in \cite{Prakash2020, Liang2020,Sasidharan2021, Krishnan2023}.  

The new proposed hierarchical  secure coded gradient aggregation problem contains $K$ users, $N$ helpers and one master.  
Each user holds a private gradient \( W_k \) and is connected to every helper, with the helpers being interconnected and all linked to the master.    
The links between users and helpers, as well as those between helpers and the master, may suffer from straggling. 
The master wishes to recover the aggregation, i.e., \( W = \sum_{k=1}^{K} W_k \), under the condition that at least \( N_r \) out of \( N \) helpers from each user are non-straggling, and at least \( N_r \) helpers  remain non-straggling in their communication with the master. 
When considering the security of a user's gradient, we account for two types of collusion:  at most  \( T \) helpers may collude with each other or with the master, and each user may collude with either the helpers or the master.     
The security constraints require that, under the two types of collusion mentioned above: (1) any set of at most \( T \) colluding helpers cannot infer any additional information about users' gradients beyond what is already known by the colluding users, and (2) the master cannot infer any additional information about users' gradients  beyond what is contained in \( W \) and what is already known by the colluding users. 

Given the parameters \( (K, N, N_r, T) \), the objective of this work is to design a hierarchical secure coded gradient aggregation scheme that minimizes the communication rate. We propose an achievable scheme  consisting of three phases, users uploading, sharing and computation, and reconstruction.  
The design of the users' upload messages is based on a globally known Vandermonde matrix, while the sharing messages between helpers are constructed using an extended Vandermonde matrix with special structural properties.    
In addition, we derive a matching converse bound, showing that when \( N_r \leq T \), secure aggregation is not feasible, whereas when \( N_r > T \), both the user-to-helper and helper-to-master optimal communication rates are \( \frac{1}{N_r - T} \).  

\subsection{Paper Organization}
The rest of this article is organized as follows. 
Section \ref{sec2} presents the model of hierarchical secure coded gradient aggregation with straggling communication links, while Section \ref{sec3} states the main results. 
Section \ref{sec4} presents a toy example and describes the proposed general hierarchical secure coded gradient aggregation scheme, while Section \ref{sec5} provides the matching converse bounds. 
Section \ref{sec6} provides the conclusion of the paper. 

\subsection{Notations}
We use $[a:b]$ to denote the set $\{a,a+1,\dots,b\}$,  and $[a]=[1:a]$ to denote the set  $\{1,2 \dots,a\}$. 
For $i_1, i_2, \cdots, i_{|\mathcal{I}|} \in  \mathcal{I}$,    
$ X_{\mathcal{I}} = (X_{i})_{i\in\mathcal{I}} = (X_{i_1},X_{i_2}, \cdots, X_{i_{|\mathcal{I}|}} ) $ denotes a vector composed of $X_i$ for each $i \in \mathcal{I}$, arranged in lexicographical order;     
$  (X_{i,j})_{i\in\mathcal{I}_j ,j\in\mathcal{J} } = (X_{\mathcal{I}_j,j} )_{ j  \in \mathcal{J}} = ( (X_{i,j})_{ i \in \mathcal{I}_j })_{  j  \in \mathcal{J} }$;   
 $ X_{\mathcal{I},\mathcal{J}}  = (X_{i,j} )_{i \in \mathcal{I}, j  \in \mathcal{J}} $;  
 $ \mathbb{F}_q $ denotes finite field with size $q$;
 $ \mathbf{0}_{m\times n} $ denotes the zero matrix with dimension $ m \times n$;
 $ \text{rank}(\mathbf{M})$ denotes the rank of matrix $\mathbf{M}$;  
$\mathbf{e}^{a}_i \in \mathbb{F}^{{a \times 1}}_q$ denotes the unit vector where the $i$-th  element  is 1 and the other elements are 0;    
$ \bot $ denotes the absence of a message being sent.



\section{System Model} \label{sec2}
\begin{figure}[htb]
	\centering 
	\includegraphics[width=0.8\linewidth]{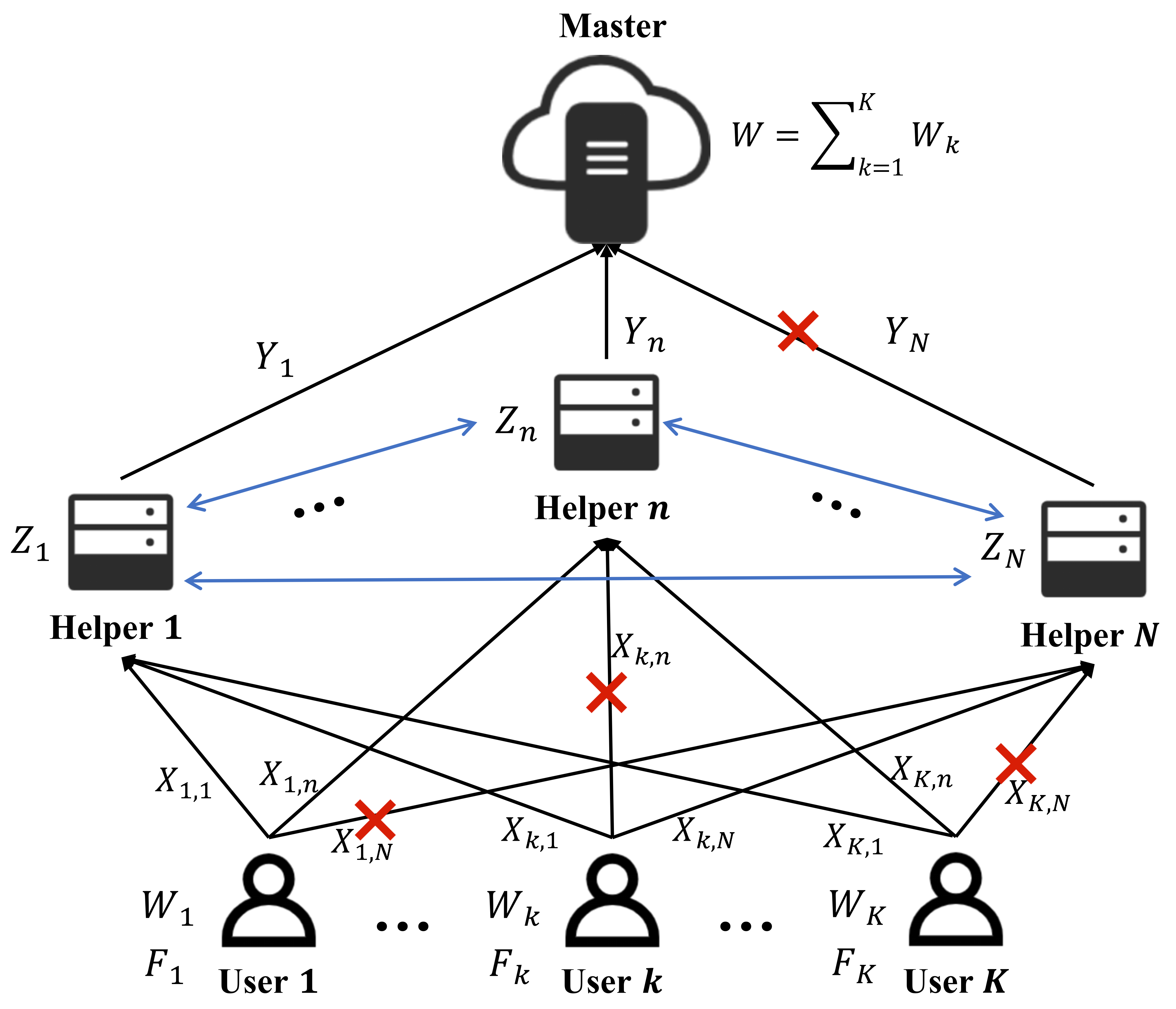}
	\caption{Hierarchical distributed learning system with straggling communication links.}\label{figsys}	
\end{figure}

Consider a hierarchical distributed learning system, which consists of \( K \) users, \( N \) helper nodes, and one master, as illustrated in Fig. \ref{figsys}. 
Each user is connected to every helper, which are interconnected among themselves, and all helpers are linked to the master.   
All communication links are secure and orthogonal. 
Partial links between users and helpers, as well as those between helpers and the master, may be straggling, while the remaining links are error-free.  
We consider the  hierarchical secure coded gradient aggregation  problem in the system described above.  
  
For each user $ k \in [K]$, let the vector $W_k$ denote  the local gradient associated with user $k$'s local data and model at one iteration.    
Each user $k$  holds the vector $W_k$  and a self-generated randomness vector $F_k$.   
Suppose the local gradients  $W_{[K]}$  and the user-side randomness vectors $F_{[K]}$ are mutually independent, and each $W_k$ is an $L \times 1$ column vector, where the $L$ elements are independent and identically distributed (i.i.d.) uniform symbols$ \footnote{ 
		  The uniformity and independence assumption for  \( W_{[K]} \) is not essential for our proposed achievable scheme, and it is only employed in the converse proof. } $  from a sufficiently large finite field  $\mathbb{F}_q$, i.e.,   
\begin{align} 
H \left( W_{[K]}, F_{[K]}\right)=	\sum_{k=1}^K H(W_k) + \sum_{k=1}^K H(F_k), \nonumber \\ 
H(W_k) = L(\text{in } q\text{-ary units}), \quad \forall k\in[K].  \label{Wk} 
\end{align} 
The master aims to calculate the summation of these local gradients, denoted as  $W  \triangleq  \sum_{k=1}^{K} W_k$, via a three-phase protocol consisting of the user uploading phase, the sharing and computation phase, and the reconstruction phase.  
\subsubsection{Users Uploading Phase}
For each $k \in [K]$ and $n \in [N]$,  user $ k $ uploads a message $X_{k,n} $ to helper $n$, which is a function of $ W_k$ and $ F_k$, and consists of $ L_X $ symbols$ \footnote{ Similar to \cite{2022SunHua}, \( L_X \) and \( L_Y \) are set as constants, where we consider the worst case, i.e., the maximum length of \( X_{k,n} \) and \( Y_n^{\boldsymbol{\nu}} \) over all possible cases.\label{foot1}} $ from the finite field $  \mathbb{F}_q$. Hence, we have   
\begin{align} \label{defXkn} 
& H \left (X_{k,n}|W_k,F_k \right )=0, \quad \forall n\in [N], k\in [K]. 
\end{align}
Due to the straggling of communication links, some helpers may fail to receive messages from user $k$.   
The \emph{user-to-helper} communication pattern can be characterized by the following vector     
\begin{align*} 
	 \boldsymbol{\nu}     \triangleq     \left( \mathcal{N}_1,\mathcal{N}_2,\dots,\mathcal{N}_K \right),  
\end{align*}
where $\mathcal{N}_k$, $k\in[K] $, denotes the set of helpers that successfully received messages from user $k$,  and $\mathcal{N}_k$ can be any subset of $[N]$ with cardinality at least  $N_r$ and $1 \leq N_r \leq N-1$.$\footnote{ 
Users who fail to upload their messages to at least $N_r$ helpers are considered stragglers in this iteration, and their messages are excluded from subsequent communications and computations.  
We assume that \(1 \leq N_r \leq N-1\); otherwise, if \(N_r = N\), the problem degenerates to a scenario where there are no straggling communication links. }$      
Thus, the set of all feasible  $\boldsymbol{\nu}$  is denoted as    $ \boldsymbol{\mathcal{N}}(N_r)  \triangleq  \big \{  \boldsymbol{\nu}  \big| \mathcal{N}_k    \subseteq [N],   |\mathcal{N}_k| \geq N_r,   k\in[K]  \big \}$.    

The helpers that do not receive any users' messages are considered stragglers and will not participate in subsequent communications and computations. The set of helpers surviving the user-to-helper communication is denoted as $ \mathcal{N}_{\text{UH}} $, i.e.,   
\begin{align*} 
	\mathcal{N}_{\text{UH}} \triangleq \bigcup_{k\in[K] }\mathcal{N}_k. 
\end{align*}  
We abuse the notation $ \mathcal{K}_n$ to denote the set of users from which helper $n$ successfully receives messages, i.e., $ \mathcal{K}_n  \triangleq  \{ k \mid n \in \mathcal{N}_k \}$.  Hence, the message received by helper $n$ from all $K$ users can be denoted by $X_{\mathcal{K}_n, n}$.

\subsubsection{Sharing and Computation Phase} 
Each helper \( n \) stores a random variable \( Z_n \), which is generated and distributed by a trusted third-party entity.
The surviving helpers, i.e., $ n \in  \mathcal{N}_{\text{UH}}$, share messages  according to the user-to-helper communication pattern.  
The message shared from helper $n$ to  helper $n'$, denoted as   $M_{n \to n'}^{ \boldsymbol{\nu}} $, is a function of $X_{\mathcal{K}_n, n}$ and $Z_n$, i.e.,    
\begin{align}  \label{defMn}
	& H\left(M^{ \boldsymbol{\nu}}_{n \to n'} \big|X_{\mathcal{K}_n, n},Z_n \right) = 0, \nonumber \\ & \quad \quad \quad \quad  \forall n,n'\in \mathcal{N}_{\text{UH}} , n \neq n',     \forall   \boldsymbol{\nu} \in 	\boldsymbol{\mathcal{N}}(N_r).   
\end{align}   
After inter-helper sharing,  each helper $n$ computes a response $Y_n^{ \boldsymbol{\nu}}$ and uploads it to the master. The response  $Y_n^{ \boldsymbol{\nu}}$ is a function of all the messages held by helper $n$
and consists $L_Y$ symbols\textsuperscript{\ref{foot1}}  from the finite field $  \mathbb{F}_q$. Hence, we have 
\begin{align} 
\label{Yn}
& H \left(Y_n^{ \boldsymbol{\nu}}   \big|  X_{\mathcal{K}_n, n},Z_n,\mathcal{M}_n^{ \boldsymbol{\nu}}  \right)=0, \nonumber \\
& \quad \quad \quad \quad \forall n \in \mathcal{N}_{\text{UH}},  \forall   \boldsymbol{\nu} \in 	\boldsymbol{\mathcal{N}}(N_r),    
\end{align}      
where  $\mathcal{M}_n^{ \boldsymbol{\nu}}  \triangleq ( M_{i \to n}^{ \boldsymbol{\nu}})_{i \in \mathcal{N}_{\text{UH}} \setminus \{n\}} $ denotes   
the messages that  helper $n$ receives from other helpers.
 
\subsubsection{Reconstruction Phase}
Some helpers may straggle in the computation  of $Y_n^{ \boldsymbol{\nu}}$ or in the  helper-to-master communication.    
Let $\mathcal{N}_{\text{HM}}$ denote the set of helpers from which the master successfully receives messages, and  $\mathcal{N}_{\text{HM}}  \subseteq \mathcal{N}_{\text{UH}}  \subseteq [N] $      
can be any set of at least $N_r$ helpers. 
The master then reconstructs the summation of the local gradients, i.e.,  $W  = \sum_{k=1}^{K} W_k $, from the received messages,  i.e., $  Y^{ \boldsymbol{\nu}}_{ \mathcal{N}_{\text{HM}}}$.

A scheme is considered achievable if it satisfies the following correctness constraint and two security constraints. 

\subsubsection*{Correctness}
The master must be able to  decode the desired summation $W$  with no error from the messages received from the helpers, i.e.,   
\begin{align} \label{correctness} 
	 & H \left (W \big|Y^{ \boldsymbol{\nu}}_{\mathcal{N}_{\text{HM}}} \right ) = 0,   \nonumber \\
	& \quad  \forall  \mathcal{N}_{\text{HM}} \subseteq  \bigcup_{k\in[K] }\mathcal{N}_k,      |\mathcal{N}_{\text{HM}}| \geq N_r, \forall   \boldsymbol{\nu} \in 	\boldsymbol{\mathcal{N}}(N_r).    
\end{align}      

Consider the security of users' local gradients  against helpers and against the master under two types of collusion: 
(1) Up to  $T$ helpers may collude with each other or with the master, where $1 \le T \le N$;$\footnote{  
We assume that \(1 \leq T \leq N\); otherwise, if \(T = 0\), it implies that the security of users' local gradients against an individual helper is not considered, which is inconsistent with the real-world scenario. }$
(2) Any user may collude with helpers or the master.  
We further impose that security constraints must be maintained, even if straggling messages are received.
More specifically, the two security constraints are described as follows. 

\subsubsection*{Security Against Helpers}  Any set of at most \( T \) colluding helpers and at most \( K \) colluding users cannot infer any additional information about \( W_{[K]} \) beyond what is already known from the colluding users, i.e.,      
\begin{align}   
	\label{security-a}  
	& I \left (  W_{[K]};  X_{[K],\mathcal{T} },    Z_{\mathcal{T}},\mathcal{M}_{\mathcal{T}}^{ \boldsymbol{\nu}}   \big| W_{\mathcal{U}}, F_{\mathcal{U}} \right) = 0,  \nonumber \\ 
	& \quad \quad   \forall  \mathcal{U} \subseteq [K],  \forall  \mathcal{T} \subseteq [N], |   \mathcal{T} | \le T,    \forall   \boldsymbol{\nu} \in 	\boldsymbol{\mathcal{N}}(N_r).      
\end{align}    

\subsubsection*{Security Against the Master}    
The master cannot infer any additional information about \( W_{[K]} \) beyond what is contained in \( W \) and the knowledge held by colluding users, even if up to \( T \) helpers and up to \( K \) users collude with the master, i.e.,  
\begin{align} 
	\label{security-b} 
	&  I \left( W_{[K]}; Y^{\boldsymbol{\nu}}_{\mathcal{N}_{\text{UH}}},  X_{[K],\mathcal{T} },  Z_{\mathcal{T}},\mathcal{M}^{\boldsymbol{\nu}}_{\mathcal{T}}  \big | W, W_{\mathcal{U}}, F_{\mathcal{U}} \right) = 0, \nonumber \\ 
	& \quad \quad \quad  \forall  \mathcal{U} \subseteq [K],  \forall  \mathcal{T} \subseteq [N], |   \mathcal{T} | \le T, \forall \boldsymbol{\nu} \in \boldsymbol{\mathcal{N}}(N_r).     
\end{align}     

We define the communication rate as the ratio that characterizes the number of symbols contained in each message relative to the number of local gradient symbols, as follows     
\begin{align}   
	R_{X} \triangleq \frac{L_X}{L}, \quad 
	R_{Y} \triangleq \frac{L_Y}{L}. \label{rate}
\end{align}     
where $R_X$, $R_Y$  denote the rates of the user-to-helper and helper-to-master communications, respectively.   
A rate tuple \( (R_X, R_Y) \) is considered achievable if there exists a scheme in which the correctness constraint \eqref{correctness} and security constraints  \eqref{security-a}  and \eqref{security-b} are satisfied, and the rates for the user-to-helper and helper-to-master communications do not exceed \( R_X \) and \( R_Y \), respectively. 
The optimal rate region, denoted as \( \mathcal{R}^* \), is defined as the closure of the set of all rate tuples that are achievable.    

\section{Main Result} \label{sec3} 
The following is the main result of this paper.
\begin{Theo}  \label{Theo1} 
	For the hierarchical secure coded gradient aggregation problem with \( K \) users, \( N  \) helpers, a resiliency threshold of \( N_r \), at most \( T \) colluding helpers, and any number of colluding users, the optimal rate region is given by 
	\begin{align}  
	\mathcal{R}^* = \left\{
	\begin{aligned}
	&\emptyset,  \quad  \quad \quad \quad  \quad \quad  \quad  \quad \quad \quad \text{when } N_r \le T  \\
	&\left\{ \left(R_X,R_Y\right) :R_X  \ge \frac{1}{N_r-T},R_Y \ge \frac{1}{N_r-T}\right\}, \nonumber \\
	&  \quad  \quad \quad \quad  \quad  \quad \quad  \quad \quad\quad \quad   \text{when } N_r > T  
	\end{aligned}  
	\right. .  
	\end{align}
\end{Theo}
\begin{IEEEproof}  
	We propose an achievable scheme for the case of \( N_r > T \) in Section \ref{sec4}, with the matching converse bounds provided in Subsections \ref{subsection2} and \ref{subsection3}, respectively. 
	In addition, for the case of \( N_r \leq T \), we show that the security constraints and the correctness constraint are contradictory, i.e., \( \mathcal{R}^* = \emptyset \), in   Subsection \ref{subsection1}.
\end{IEEEproof}

\begin{remark}	 
	 From Theorem \ref{Theo1}, we can observe that when the resiliency threshold does not exceed the number of potentially colluding users, i.e., \( N_r \leq T \), there exists no achievable scheme for the hierarchical secure coded gradient aggregation problem defined in Section \ref{sec2}.   
	 Intuitively, if all servers surviving the helper-to-master communication collude, they could derive the desired summation \( W \), thereby violating the security against helpers constraint.   
\end{remark}

\begin{remark}	 
   When  \( N_r < T \),  the optimal rates of the user-to-helper and helper-to-master communications depend on the difference between the resiliency threshold \( N_r \) and the number of  potentially colluding helpers \( T \).  
	When \( N_r \) increases, fewer symbols need to be transmitted, whereas a larger \( T \) requires more symbols to be transmitted.    
\end{remark}   
 
\section{Proof of  Theorem \ref{Theo1}: Achievability when $N_r > T$} \label{sec4}
In this section, we provide the achievability proof of Theorem \ref{Theo1}. We begin by using a simple example to illustrate the idea of the new proposed scheme.    
\subsection{Motivating Example: $K=2,N=4,N_r = 3,T=1$} 
Consider $K = 2$ users and  $N = 4$ helpers,  where at least $N_r = 3$ helpers successfully  receive the message from user $k$.    
Suppose the message length is $L = N_r-T = 2$, and the parameter of the large enough finite field is $q = 7$, i.e., $W_k = \left[ W_{k,1}, W_{k,2}  \right]^{\mathsf{T}} \in \mathbb{F}_7^{2 \times 1} $.   
We describe the achievable scheme under the communication pattern $\boldsymbol{\nu} =   \left(  \mathcal{N}_1, \mathcal{N}_2  \right) = \left(\{1,2,3\}, \{1,2,4\} \right)$, $ \mathcal{N}_{\text{HM}} =  \{2,3,4\}$.  

In the \emph{users uploading phase}, each user uses a globally known Vandermonde matrix, to mix the local gradient with a self-generated randomness, and then uploads the encoded messages to different helpers. More specifically, the message that user $k$ uploads to helper node $n$, i.e.,  $X_{k,n}$,  is set as    
\begin{align} \label{exD}   
	\begin{bmatrix} X_{k,1} \\X_{k,2}\\X_{k,3}\\X_{k,4} \end{bmatrix} 
	=\mathbf{V}  
	\begin{bmatrix} W_{k,1}\\W_{k,2}\\F_k\end{bmatrix}  
	=\begin{bmatrix} 1& 1& 1^2 \\ 1&	2&	2^2\\1&	3&	3^2 \\1&	4&	4^2\end{bmatrix}
	\begin{bmatrix} W_{k,1}\\W_{k,2}\\F_k\end{bmatrix}, \quad k = 1,2,  
\end{align}  
where $F_1$ and $F_2$ are $2$ i.i.d. uniform symbols over \( \mathbb{F}_7 \), and  $\mathbf{V}$ is a globally known Vandermonde matrix. 
 
Each helper stores a decoding matrix for computation, denoted as \( \mathbf{S}_n \), and a helper-side randomness, denoted by  \( Z_n \), generated and distributed by a trusted third-party entity.  For any $n \in [4]$,  the decoding matrix \( \mathbf{S}_n \) is set as
\begin{align*}
     \mathbf{S}_n = \mathbf{V}\mathbf{G}_n^{-1}, \text{ where  } \mathbf{G}_n  =   \begin{bmatrix}
    	1& n & n^2 \\  
    	1& 5 & 5^2 \\ 
    	1& 6 & 6^2 \\ 
    \end{bmatrix}.
\end{align*}      
Take \( n = 3, 4 \) as examples, we have
\begin{align*} 
	\mathbf{S} _3  =  \begin{bmatrix} 1& 2& 5 \\ 2& 5& 1\\ 1& 0& 0 \\ 5 & 1& 2 \end{bmatrix},   \text{ and } 
	\mathbf{S} _4  =  \begin{bmatrix} 3& 6& 6 \\ 6& 6& 3\\ 3& 4& 1 \\ 1 & 0& 0 \end{bmatrix}.
\end{align*}   
Consider $16$ i.i.d. uniform symbols over $\mathbb{F}_7$, denoted as   $Q_{n,j}^{(k)}, n\in[4],j\in[2], k \in [2]$.  Then, for any $i \in [4]$, the helper-side randomness $Z_i$ is set as     
\begin{align*}  
	& Z_{i} =   \left(Z_{i,n}^{(k)} \right)_{ n \in[4]\setminus{\{i\}},k \in [2]},  \text{ where}    \nonumber \\
		 &   \begin{bmatrix} Z_{1,n}^{(k)} \\Z_{2,n}^{(k)} \\ Z_{3,n}^{(k)} \\Z_{4,n}^{(k)}\end{bmatrix} =  \mathbf{S}_n      \begin{bmatrix} 
		0& 0  \\
		1& 5 \\  
		1& 6 \\ 
	\end{bmatrix}  
	\begin{bmatrix} Q_{n,1}^{(k)} \\Q_{n,2}^{(k)} \end{bmatrix}. 
\end{align*}        
Take \( n = 3, 4 \) as examples, we have  
\begin{align} \label{ex_z} 
\begin{bmatrix} Z_{1,3}^{(k)} \\Z_{2,3}^{(k)} \\ Z_{3,3}^{(k)} \\Z_{4,3}^{(k)}\end{bmatrix} = 
\begin{bmatrix} 0 &	5 \\
	6 &	3 \\
	0 &	0 \\
	3 &	3   \end{bmatrix}
\begin{bmatrix} Q_{3,1}^{(k)} \\Q_{3,2}^{(k)} \end{bmatrix}, \quad 
\begin{bmatrix} Z_{1,4}^{(k)}\\Z_{2,4}^{(k)}\\Z_{3,4}^{(k)} \\Z_{4,4}^{(k)} \end{bmatrix} = 
\begin{bmatrix} 5&	3 \\ 
	2&	6 \\ 
	5&	5 \\ 
	0&	0 
 \end{bmatrix}
\begin{bmatrix} Q_{4,1}^{(k)} \\Q_{4,2}^{(k)} \end{bmatrix}.
\end{align}   

In the \emph{sharing and computation phase}, based on the pattern of user-to-helper  communication, the messages between helpers are set as follows  
\begin{align} \label{ex4_M} 
	M_{i \to j}^{\boldsymbol{\nu} } = \left\{  
	\begin{aligned}
		&  X_{2,i}+Z_{i,3}^{(2)},\quad   i\in \mathcal{N}_2,j=3  \\
		&  X_{1,i}+Z_{i,4}^{(1)},\quad   i\in \mathcal{N}_1,j=4  \\
		& \bot , \quad  \quad \quad  \quad \quad     i\in[4]\setminus \{j\},j = 1,2  
	\end{aligned} 
	\right. .  
\end{align}
Thus,  we have $ \mathcal{M}^{\boldsymbol{\nu}}_3  = \left( M_{1 \to 3}^{\boldsymbol{\nu}}, M_{2 \to 3}^{\boldsymbol{\nu}},M_{4 \to 3}^{\boldsymbol{\nu}}  \right) $, $ \mathcal{M}^{\boldsymbol{\nu}}_4  = \left(  M_{1 \to 4}^{\boldsymbol{\nu}}, M_{2 \to  4}^{\boldsymbol{\nu}},M_{3 \to 4}^{\boldsymbol{\nu}}   \right)$, and $ \mathcal{M}^{\boldsymbol{\nu}}_1 =  \mathcal{M}^{\boldsymbol{\nu}}_2  =  \bot$.  

Then, helper $3$ recovers the messages it failed to receive from $\mathcal{M}^{\boldsymbol{\nu}}_3$ by using the inverse matrix of the submatrix of $\mathbf{S}_3$ as the decoding matrix, i.e.,   
\begin{align}  
& \hat{X}_{2,3}  
 =   \left[  \mathbf{e}_1^{3} \right]^{\mathsf{T}}  \left( \left[   (  \mathbf{e}_i^{4})_{ i \in   \mathcal{N}_2  }    \right]^{\mathsf{T}}    \mathbf{S}_3      \right)^{-1}  \left[ M_{1 \to 3}^{\boldsymbol{\nu}}, M_{2 \to  3}^{\boldsymbol{\nu}},M_{4 \to 3}^{\boldsymbol{\nu}}     \right]^{\mathsf{T}}.       \nonumber  
\end{align}  
Similarly, helper $4$ recovers ${X}_{1,4}$ by following  
\begin{align*}   
  \hat{X}_{1,4}  
	=   \left[  \mathbf{e}_1^{3} \right]^{\mathsf{T}}  \left( \left[   (  \mathbf{e}_i^{4})_{ i \in   \mathcal{N}_1  }    \right]^{\mathsf{T}}    \mathbf{S}_4      \right)^{-1}  \left[ M_{1 \to 4}^{\boldsymbol{\nu}}, M_{2 \to 4}^{\boldsymbol{\nu}},M_{3 \to 4}^{\boldsymbol{\nu}}     \right]^{\mathsf{T}}.       
\end{align*}  
The message that helper $n$ uploads to the master, i.e.,  $Y_n^{\boldsymbol{\nu} } $,  is set as      
\begin{align}\label{ex4_Y}
	Y_n^{\boldsymbol{\nu} }  = \left\{   
	\begin{aligned}
		&  X_{1,n} + X_{2,n},  \quad n = 1,2 \\
		&  X_{1,3} + \hat{X}_{2,3} , \quad n=3  \\
		&  \hat{X}_{1,4} + {X}_{2,4} , \quad  n=4 
	\end{aligned}
	\right. .   
\end{align}  

\subsubsection*{Proof of Correctness}
We show the correctness of the above scheme as follows.  From the definition of $ \hat{X}_{2,3} $, we have 
\begin{subequations}  
	\begin{align}   
		& \hat{X}_{2,3}  
		=   \left[  \mathbf{e}_1^{3} \right]^{\mathsf{T}}  \left( \left[   (  \mathbf{e}_i^{4})_{ i \in   \mathcal{N}_2  }    \right]^{\mathsf{T}}    \mathbf{S}_3      \right)^{-1}  \left[ M_{1 \to 3}^{\boldsymbol{\nu}}, M_{2 \to 3}^{\boldsymbol{\nu}},M_{4 \to 3}^{\boldsymbol{\nu}}     \right]^{\mathsf{T}}       \nonumber \\ 
		= &  \left[  \mathbf{e}_1^{3} \right]^{\mathsf{T}} \begin{bmatrix} 1& 2& 5 \\ 2& 5& 1 \\5 & 1& 2 \end{bmatrix}^{-1}     \begin{bmatrix} X_{2,1}+Z_{1,3}^{(2)}  \\ X_{2,2}+Z_{2,3}^{(2)}   \\ X_{2,4}+Z_{4,3}^{(2)}  \end{bmatrix}   \label{ex23a} \\ 
		\label{ex23b}
		= & \left[  \mathbf{e}_1^{3} \right]^{\mathsf{T}}   \begin{bmatrix}    2  & 1 &  5 \\ 1 &  5 &  2 \\ 5  & 2  & 1 \end{bmatrix}   	
		\begin{bmatrix} 1& 1& 1& 0& 5 \\1& 2& 2^2& 6& 3\\ 1& 4& 4^2& 3& 3 \end{bmatrix}   
		\begin{bmatrix} W_{2,1}  \\ W_{2,2}  \\F_2 \\  Q_{3,1}^{(2)} \\  Q_{3,2}^{(2)}   \end{bmatrix}    \\ 
		= &      W_{2,1} +    3 W_{2,2} +   2   F_2  = {X}_{2,3},   \label{ex23c}
	\end{align}  
\end{subequations}    
where  \eqref{ex23a} follows from the design of $M_{i \to j}^{\boldsymbol{\nu} } $, i.e., \eqref{ex4_M}, 
\eqref{ex23b} follows from   \eqref{exD} and \eqref{ex_z},   
and  \eqref{ex23c} follows from the design of $X_{2,3}$ in \eqref{exD}. 

Similarly, we have  
	\begin{align}  
		& \hat{X}_{1,4}   
		=    \left[  \mathbf{e}_1^{3} \right]^{\mathsf{T}} \begin{bmatrix} 3& 6& 6 \\ 6& 6& 3 \\3 & 4& 1 \end{bmatrix}^{-1}     \begin{bmatrix} X_{1,1}+Z_{1,4}^{(1)}  \\ X_{1,2}+Z_{2,4}^{(1)}   \\ X_{1,3}+Z_{4,4}^{(1)}  \end{bmatrix}     \nonumber \\ 
		= & \left[  \mathbf{e}_1^{3} \right]^{\mathsf{T}}   \begin{bmatrix}    	1  &  4  &  3 \\  3 &  6 &  6 \\
			6 &  6 &  3 
		  \end{bmatrix}   	
		\begin{bmatrix}   1  &  1  &  1  &  5  &  3 \\ 
			1  &  2  &  2^2  &  2  &  6 \\
			1   & 3  &  3^2  & 5  &   5  \end{bmatrix}   
		\begin{bmatrix} W_{1,1}  \\ W_{1,2}  \\F_1 \\  Q_{4,1}^{(1)} \\  Q_{4,2}^{(1)}   \end{bmatrix}  \nonumber  \\ 
		= &      W_{1,1} +    4 W_{1,2} +   2   F_1  = {X}_{1,4}.   \label{ex23d}  
	\end{align}    
Substituting the result of \eqref{ex23c} and  \eqref{ex23d} into \eqref{ex4_Y}, we have  $Y_n  = X_{1,n} + X_{2,n}, \forall n \in [4]$. 
As any 3-row submatrix of \(\mathbf{V}\) retains the structure of a Vandermonde matrix,  
we can recover \(W_{1,1} + W_{2,1}\), \(W_{1,2} + W_{2,2}\), and \(F_1 + F_2\) from the messages \((X_{1,n} + X_{2,n})_{n \in \mathcal{N}_{\text{HM}} =  \{2,3,4\} }\).    Then, we can obtain \( W = W_1 + W_2 = (W_{1,1} + W_{2,1}, W_{1,2} + W_{2,2}) \).

When considering security against helpers or the master, we provide a security proof using \(\mathcal{T} = \{3\}\) and \(\mathcal{U} = \emptyset\) as an example to demonstrate how the scheme ensures security.   
The security proofs in other cases follow similar steps and can be found in the general proof in Subsection \ref{secgeneral}.    
\subsubsection*{Security Againist Helpers}   
Suppose  \(\mathcal{T} = \{3\}\) and \(\mathcal{U} = \emptyset\).  Then, following the message design shown above, we have    
\begin{subequations} 
\begin{align}   
&	I \left (  W_1,W_2 ;  X_{1,3}, X_{2,3}, \mathcal{M}_{3}^{ \boldsymbol{\nu}} ,  Z_{3}  \right) \nonumber \\
=	& I \Big (  W_1,W_2 ;  X_{1,3},  X_{2,1}+ 5Q_{3,2}^{(2)},       \nonumber \\ & \quad \quad   X_{2,2}+ 6Q_{3,1}^{(2)}+3Q_{3,2}^{(2)},   X_{2,4}+3Q_{3,1}^{(2)}+3Q_{3,2}^{(2)}     \Big) \label{exa}  \\ 
= &  H  \Big ( X_{1,3},   X_{2,1}+ 5Q_{3,2}^{(2)},    X_{2,2}+ 6Q_{3,1}^{(2)}+3Q_{3,2}^{(2)},\nonumber \\ & \quad     X_{2,4}+3Q_{3,1}^{(2)}+3Q_{3,2}^{(2)}     \Big) -  H \Big(F_1,   F_2 +  5Q_{3,2}^{(2)},   \nonumber \\    
&  4F_2 + 6Q_{3,1}^{(2)}+3Q_{3,2}^{(2)}, 2F_2 +  3Q_{3,1}^{(2)}+3Q_{3,2}^{(2)}  \Big| W_1,W_2 \Big)  \label{exb}  \\  
\le &  4 - H \Big(F_1,   F_2,  Q_{3,1}^{(2)},   Q_{3,2}^{(2)} \Big| W_1,W_2 \Big)   \label{exc}  \\
 = &  4 -  4 =  0, \label{exd} 
\end{align}      
\end{subequations}   
where \eqref{exa} follows from the fact that $X_{2,3}$ can be decoded from $\mathcal{M}_{3}^{ \boldsymbol{\nu}}$ (see  \eqref{ex23c}), and the fact that $Z_3$ is a function of $ \left( Q_{i,j}^{(k)}\right)_{i\in \{1,2,4\}, j \in \{1,2\}, k\in\{1,2\} } $, which is independent of $\left(W_1,W_2,X_{1,3},X_{2,3},\mathcal{M}_{3}^{ \boldsymbol{\nu}} \right)$,  
\eqref{exb} follows from \eqref{exD}, the first term of \eqref{exc} follows from the fact that $X_{1,3}, M_{1 \to 3}^{\boldsymbol{\nu}}, M_{2 \to 3}^{\boldsymbol{\nu}},M_{4 \to 3}^{\boldsymbol{\nu}}$  each contain  $1$ symbol,  
the second term of \eqref{exc} follows from the fact that  $F_2 +  5Q_{3,2}^{(2)},     4F_2 + 6Q_{3,1}^{(2)}+3Q_{3,2}^{(2)}, 2F_2 +  3Q_{3,1}^{(2)}+3Q_{3,2}^{(2)}$ are $3$   
linearly independent linear combinations of \(F_2,Q_{3,1}^{(2)}, Q_{3,2}^{(2)}\),   
and   \eqref{exd}  follows from  the fact that  
$F_1, F_2,  Q_{3,1}^{(2)}, Q_{3,2}^{(2)}$ are $4$ i.i.d. uniform   random variables, and are independent of $W_1,W_2$.  
Since mutual information is non-negative, \eqref{exd} implies that  the security condition against helpers, i.e., \eqref{security-a}, holds for \(\mathcal{T} = \{3\}\) and \(\mathcal{U} = \emptyset\).

\subsubsection*{Security Againist the Master}  
First, we note that the messages received by the master can be determined by $W_1 + W_2, X_{1,3},X_{2,3}$, i.e.,   
\begin{subequations}  
\begin{align}   
	& H \left(   Y^{\boldsymbol{\nu}}_{[4]} \Big|W_1 + W_2, X_{1, 3},X_{2, 3} \right)  \nonumber \\ 
	 = &   H \left(  \left( X_{1,n} + X_{2,n} \right)_{n \in [4]}  \Big |W_1 + W_2,   X_{1, 3},X_{2, 3}  \right)   \label{ex2a}  \\  
	 = &   H \left(  W_{1,1} + W_{2,1}, W_{1,2} + W_{2,2},  F_1 + F_2 \big |W_1 + W_2, \right. \nonumber \\
	 & \left. \quad W_{1,1} + 3 W_{1,2} + 2F_1 ,W_{2,1} + 3 W_{2,2} + 2F_2  \right)   \label{ex2b}  \\  
	 \le  &   H \left(   F_1 + F_2 \big |W_1 + W_2,  F_1+F_2  \right)     = 0,   \label{ex2c}
\end{align} 
\end{subequations}
where \eqref{ex2a} follows from the proof of correctness, 
and \eqref{ex2b} follows from \eqref{exD}. 

Thus, suppose  \(\mathcal{T} = \{3\}\) and \(\mathcal{U} = \emptyset\), then we have    
\begin{subequations} 
	\begin{align}  
	&	I \left (  W_1,W_2 ;  Y^{\boldsymbol{\nu}}_{[4]}, X_{1,3}, X_{2,3},    \mathcal{M}_{3}^{ \boldsymbol{\nu}} ,  Z_{3} \Big |  W_1 + W_2    \right) \nonumber \\ 
	\le  &  I \left (  W_1,W_2 ;    X_{1,3}, X_{2,3},    \mathcal{M}_{3}^{ \boldsymbol{\nu}} ,  Z_{3} \Big |  W_1 + W_2    \right)   \label{ex3a}   \\
	\le  &  I \left (  W_1,W_2 ;    X_{1,3}, X_{2,3},    \mathcal{M}_{3}^{ \boldsymbol{\nu}} ,  Z_{3}    \right)   \label{ex3b}  \\
	\le  &  0  \label{ex3c},    
	\end{align}      
\end{subequations}  
where \eqref{ex3a} follows from \eqref{ex2c},  \eqref{ex3b} follows from  the fact that $W_1 + W_2$ is a function of $W_1,W_2$, and   \eqref{ex3c} follows from the result of \eqref{exd}.
Since mutual information is non-negative, \eqref{ex3c} implies that  the security  against the master, i.e., \eqref{security-b}, holds for the case \(\mathcal{T} = \{3\}\) and \(\mathcal{U} = \emptyset\).

\subsubsection*{Rate}  As $L_X  = L_Y = 1$, we have $R_X = R_Y = \frac{1}{L} = \frac{1}{2}$.  
\subsection{General Proof for Arbitrary $K,N,N_r,T$} \label{secgeneral}
We now generalize the above example to arbitrary values of \( K, N, N_r  \) and \( T \) in order to present the achievability proof for Theorem \ref{Theo1}.   
Since \( \mathcal{R}^* = \emptyset \) when \( N_r \le T \), it suffices to consider scenarios where \( N_r > T \).   
We assume that the size of the finite field, \( q \), is sufficiently large, i.e., \( q \ge N + N_r \).  
For cases where the finite field size is insufficient, a method similar to that discussed in \cite{2022SunHua} can be applied to obtain an extended field size. We omit this discussion here.

\emph{Users Uploading Phase.}  
Each  user $k$, $k \in [K]$,  independently generates a uniform $\frac{LT}{N_r-T}  \times 1$ column vector over $\mathbb{F}_q$ and assigns this vector to $F_k$. 
Then, user $k$ divides each message $W_k$ into $N_r-T$ parts and each user-side randomness $F_k$ into $T$ parts, i.e.,    
\begin{align*} 
W_k = &  [W_{k,1}^{\mathsf{T}},W_{k,2}^{\mathsf{T}}, \dots, W_{k,N_r-T}^{\mathsf{T}} ]^{\mathsf{T}} , \nonumber \\
F_k = &  [ F_{k,1}^{\mathsf{T}},F_{k,2}^{\mathsf{T}}, \dots,  F_{k,T}^{\mathsf{T}} ]^{\mathsf{T}} . 
\end{align*} 
where each $W_{k,i}$, $i\in[N_r-T]$, and $F_{k,j}$, $j \in[T]$, has  $l = \frac{L}{N_r-T}$ symbols.   
The message that user $k$ uploads to helper node $n$, i.e.,  $X_{k,n}$,  is set as  
\begin{align} 
	& \mathbf{V}   =  \begin{bmatrix}
		1& \alpha_1  & \cdots & \alpha_1^{N_r-1} \\
		1& \alpha_2  & \cdots & \alpha_2^{N_r-1} \\
		\vdots & \vdots  & \ddots  & \vdots \\ 
		1& \alpha_{N}  & \cdots & \alpha_{N}^{N_r-1} \\
	\end{bmatrix}, \text{ and} \label{Verdo} \\   
	& [X_{k,1}, X_{k,2}, \dots, X_{k,N}]^{\mathsf{T}}   \nonumber  \\
	= &  \mathbf{V}  [W_{k,1}, W_{k,2}, \dots, W_{k,N_r-T}, F_{k,1}, F_{k,2}, \dots, F_{k,T} ]^{\mathsf{T}},  \label{X_kn}
\end{align}   
where $\alpha_i \neq 0$, $i\in[N]$,  differs from each other over  $\mathbb{F}_q$. 
 
\emph{Sharing and Computation Phase.}
For $ n \in[N] $, we set 
\begin{align} \label{gen_G}
\mathbf{G}_n = &  \left[ 
\begin{array}{cccc}
1& \alpha_n  & \cdots & \alpha_n^{N_r-1} \\
\hdashline
1& \alpha_{N+1}  & \cdots & \alpha_{N+1}^{N_r-1} \\
1& \alpha_{N+2}  & \cdots & \alpha_{N+2}^{N_r-1} \\
\vdots & \vdots  & \ddots  & \vdots \\ 
1& \alpha_{N+N_r-1}  & \cdots & \alpha_{N+N_r-1}^{N_r-1}  \\
\end{array}
\right],  \\  \label{gen_G2} 
\overset{\sim}{\mathbf{G}}    = & \left[ 
\begin{array}{cccc}
	0& 0  & \cdots & 0 \\
	\hdashline
	1& \alpha_{N+1}  & \cdots & \alpha_{N+1}^{N_r-2} \\
	1& \alpha_{N+2}  & \cdots & \alpha_{N+2}^{N_r-2} \\
	\vdots & \vdots  & \ddots  & \vdots \\ 
	1& \alpha_{N+N_r-1}  & \cdots & \alpha_{N+N_r-1}^{N_r-2}  \\
\end{array} 
\right],   
\end{align}  
where $\alpha_i \neq 0$, $i\in[N+N_r-1]$,  differs from each other over  $\mathbb{F}_q$, and  $\alpha_i, i\in[N]$, are the same as those used in  \eqref{Verdo}. 
\begin{figure*} 
	\begin{align}
		& \mathbf{S}_n \left[\mathbf{G}_n, \overset{\sim}{\mathbf{G}} \right] 	 
		\left[W_{k,1}, W_{k,2}, \cdots, W_{k,N_r-T},    
		F_{k,1}, F_{k,2},   \cdots,   F_{k,T}, 
		Q_{n,1}^{(k)}, Q_{n,2}^{(k)},   \cdots,  Q_{n,N_r-1}^{(k)}   
		\right]^{\mathsf{T}} \nonumber \\
		=   &  \mathbf{V} \left[W_{k,1}, W_{k,2}, \dots, W_{k,N_r-T}, F_{k,1}, F_{k,2}, \dots, F_{k,T} \right]^{\mathsf{T}}  + 
		\mathbf{S}_n   \overset{\sim}{\mathbf{G}}   
		\left[ 
		Q_{n,1}^{(k)}, Q_{n,2}^{(k)},   \cdots,  Q_{n,N_r-1}^{(k)}  
		\right]^{\mathsf{T}}  \nonumber \\
		=   &  \left[X_{k,1}, X_{k,2}, \dots, X_{k,N} \right]^{\mathsf{T}}  
		+ \left[ Z_{1 , n}^{(k)},Z_{2, n}^{(k)}, \dots, Z_{N, n}^{(k)} \right]^{\mathsf{T}}   
		= \left[M_{1, n}^{(k)},M_{2, n}^{(k)}, \dots, M_{N, n}^{(k)}\right]^{\mathsf{T}}  \label{design}.    
	\end{align}
\end{figure*}  
Consider $NK(N_r-1)$ i.i.d. uniform $l  \times 1$ vectors over $\mathbb{F}_q^{l \times 1}$, denoted as $Q_{i,j}^{(k)},  i \in[N], j \in[N_r-1], k \in [K]$.  
Then, for \( i \in [N] \), \( n \in [N] \) and \( k \in [K] \), we define \( Z_{i, n}^{(k)} \) and \( M_{i, n}^{(k)} \), which satisfy the equation in \eqref{design}. Their explicit definitions are as follows.   

As $\mathbf{G}_n$ is a Vandermonde matrix of dimension $N_r \times N_r$, it is of full rank and invertible. Consequently, we can compute 
\begin{align} \label{gen_S} 
\mathbf{S}_n = \mathbf{V} \mathbf{G}_n^{-1}.
\end{align}
The helper-side randomness of helper $i$, where $i \in [N]$, is given by    
\begin{align}
	Z_{i,n}^{(k)}= &  \left(  \mathbf{S}_n    \overset{\sim}{\mathbf{G}}  
	\left[ 
	Q_{n,1}^{(k)}, Q_{n,2}^{(k)},   \cdots,  Q_{n,N_r-1}^{(k)}  
	\right]^{\mathsf{T}}   \right)^{\mathsf{T}} \mathbf{e}_i ^{N}    ,  \nonumber  \\
	& \forall n\in[N], i\in[N], k\in[K], \text{  and} \label{ach_Zn}   \nonumber  \\
	Z_{i} =  & \left(Z_{i,n}^{(k)} \right)_{ n \in[N]\setminus{\{i\}},k \in [K]}, \quad \forall i \in[N].
\end{align}      
\begin{remark} \label{remarkZ}
	From the definition of $\mathbf{S}_n$, i.e., \eqref{gen_S}, we note that the first element of the $n$-th row of $\mathbf{S}_n$ is $1$, and all other elements are $0$, i.e., 
	\begin{align*}
	 \left( 	\mathbf{e}_n^{N}  \right)^{\mathsf{T}} \mathbf{S}_n  = & (1,\mathbf{0}_{1\times (N_r-1)}).   
	\end{align*}
	As a result,  we note that for any $k \in [K]$ and any $ n \in [N]$, 
	\begin{align*}
		\left(    Z_{n,n}^{(k)}  \right)^{\mathsf{T}} = &  \left(  	\mathbf{e}_n^{N}   \right) ^{\mathsf{T}}     \mathbf{S}_n    \overset{\sim}{\mathbf{G}}  
		\left[  
		Q_{n,1}^{(k)}, Q_{n,2}^{(k)},   \cdots,  Q_{n,N_r-1}^{(k)}    
		\right]^{\mathsf{T}} \\
		= & \mathbf{0}_{1\times l}.   
	\end{align*}    
\end{remark}

The inter-helper message  $M_{n,i}^{(k)}$, transmitted from helper $n$ to helper $i$ in service of user $k$, is given by 
\begin{align}
	\label{ach_M0} 
M_{n,i}^{(k)} =  X_{k,n} + Z_{n,i}^{(k)}, \quad \forall i \in  [N] \setminus \{n\},  n \in [N],  k \in [K]. 
\end{align} 
Based on the pattern of the \emph{user-to-helper}  communication,  i.e., \( \boldsymbol{\nu} \), for surviving helpers, i.e., \( n \in \mathcal{N}_{\text{UH}} \), we define     
\begin{align}  
M_{n \to i}^{\boldsymbol{\nu}} = & \left(M_{n,i}^{(k)}  \right)_{k\in \mathcal{K}_n \cap([K]\setminus\mathcal{K}_i)}, \nonumber \\
&  \quad \quad \forall i \in [N] \setminus \{n\}, n \in \mathcal{N}_{\text{UH}}. 
\label{ach_M}   
\end{align}   
Then, the helper \( n \),  \( n \in \mathcal{N}_{\text{UH}} \) computes \( \hat{X}_{k,n} \), \( k \in [K] \setminus \mathcal{K}_n \),   based on the received messages $\mathcal{M}_{n}^{\boldsymbol{\nu}} = \left( M_{i \to n}^{\boldsymbol{\nu}}  \right)_
{  i \in \mathcal{N}_{\text{UH}} \setminus \{n\}} $, by following  
\begin{align}  \label{hatXkn}
	 \hat{X}_{k,n}  = &  \left[ 	\left(M_{i,n}^{(k)} \right)_{ i \in \mathcal{N}'_k   }   \right]   \left(\mathbf{S}_n^{\mathsf{T}} \left[   \left(   \mathbf{e}_i^{N} \right)_{ i \in \mathcal{N}'_k   }     \right]   \right)^{-1} \mathbf{e}_1^{N_r},         
\end{align}   
where $\mathcal{N}_{k}'$ can be any helper set such that $\mathcal{N}_k'  \subseteq \mathcal{N}_k, |\mathcal{N}_k'| = N_r$.     
\begin{remark}  
	From the definition  of $\mathbf{S}_n$, i.e., \eqref{gen_S},   
	we note that for any   set  $\mathcal{N} \subseteq [N]$ such that $|\mathcal{N}| \le  N_r$,  
	\begin{align} \label{rank0}
		& \text{rank} \left(  \left[ \left( \mathbf{e}_i^{N} \right)_{ i \in \mathcal{N}   }   \right] ^{\mathsf{T}}     \mathbf{S}_n    \right)   \nonumber \\ 
		=  &  \text{ rank} \left(  \left[ \left( \mathbf{e}_i^{N} \right)_{ i \in \mathcal{N}   }   \right] ^{\mathsf{T}}      \mathbf{V}  \mathbf{G}_n^{-1}     \right)    =  |\mathcal{N} |,     
	\end{align} 
	where 	\eqref{rank0} follows from the fact that \(\mathbf{G}_n\) is an invertible \(N_r \times N_r\) matrix, and \(\mathbf{V}\) is a Vandermonde matrix of dimensions \(N \times N_r\).  
	Thus,  we note that   
	$  \text{rank} \left(  \left[ \left( \mathbf{e}_i^{N} \right)_{ i \in \mathcal{N}_k'  }   \right] ^{\mathsf{T}}     \mathbf{S}_n    \right)      = N_r$, which implies that   $	 \left[ \left( \mathbf{e}_i^{N} \right)_{ i \in \mathcal{N}_k'  }   \right] ^{\mathsf{T}}     \mathbf{S}_n    $ is invertible.
\end{remark}   

Finally, we set  
\begin{align}  \label{Y_n}
	Y_n^{ \boldsymbol{\nu} }  = \sum_{k\in\mathcal{K}_n}   X_{k,n} + \sum_{k\in[K] \setminus \mathcal{K}_n } \hat{X}_{k,n},  \quad  \forall n\in\mathcal{N}_{\text{UH}}.
\end{align}

\subsubsection*{Proof of Correctness}  
Based on the calculation of $\hat{X}_{k,n}$  (see \eqref{hatXkn}), for any $k \in [K]$ and $n \in [N]$, we have       
\begin{subequations}
\begin{align} 
	&  \hat{X}_{k,n}^{\mathsf{T}}    
	=  
	  \left[ \mathbf{e}_1 ^{N_r} \right]^{\mathsf{T}}     \left(  \left[ \left( \mathbf{e}_i^{N}\right)_{ i \in \mathcal{N}'_k} \right]^{\mathsf{T}} \mathbf{S}_n    \right)^{-1}       \left[ 	\left(M_{i, n}^{(k)} \right)_{ i \in \mathcal{N}'_k   }   \right]^{\mathsf{T}} \nonumber  \\ 
	= 
	 \label{AhatXkneq}
	 &  \left[ \mathbf{e}_1 ^{N_r} \right]^{\mathsf{T}}     
	 \left(  \left[  \left( \mathbf{e}_i^{N}\right)_{ i \in \mathcal{N}'_k} \right]^{\mathsf{T}} \mathbf{S}_n    \right)^{-1}   \nonumber \\ 
	 &   \Bigg(   \left[ \left ( \mathbf{e}_i^{N} \right )_{ i \in \mathcal{N}'_k} \right]^{\mathsf{T}} \mathbf{S}_n    \left[\mathbf{G}_n, \overset{\sim}{\mathbf{G}} \right] 	     \nonumber  \\ 
	 &       \left[ \left(W_{k,i}\right)_{ i\in [N_r-T] },    \left(F_{k,i}\right)_{ i\in [T] },  \left(Q_{n,i}^{(k)} \right)_{ i\in [N_r-1] } 
	 \right]^{\mathsf{T}} \Bigg)  \\
	 =    & \left[ \mathbf{e}_1 ^{N_r} \right]^{\mathsf{T}}      
	\left[\mathbf{G}_n, \overset{\sim}{\mathbf{G}} \right] 	   \nonumber \\ 
	 & \quad     \left[ \left(W_{k,i}\right)_{ i\in [N_r-T] },    \left(F_{k,i}\right)_{ i\in [T] },  \left(Q_{n,i}^{(k)} \right)_{ i\in [N_r-1] } 
	 \right]^{\mathsf{T}} \nonumber    \\ 
	  \label{ChatXkneq}
	 = 
	  &    \sum_{i=1}^{N_r-T} \alpha_n^{i-1} W_{k,i}^{\mathsf{T}}    + \sum_{i=N_r-T+1}^{N_r} \alpha_n^{i-1} F_{k,i-N_r+T}^{\mathsf{T}}   \\
	  = 
	  &    {X}_{k,n}^{\mathsf{T}}    \label{DhatXkneq}   , 
\end{align} 
\end{subequations}   
where \eqref{AhatXkneq}  follows from \eqref{design},
\eqref{ChatXkneq}   follows from \eqref{gen_G} and \eqref{gen_G2}, 
and \eqref{DhatXkneq}     follows from \eqref{Verdo} and \eqref{X_kn}. 

Substituting the result of \eqref{DhatXkneq} into \eqref{Y_n}, we have  
\begin{align} \label{correctness2}
	 Y_n^{\boldsymbol{\nu}} = \sum_{k=1}^{K}  X_{k,n}, \quad \forall n\in \mathcal{N}_{\text{UH}}.   
\end{align}
For any $\mathcal{N}_{\text{HM}} \subseteq \mathcal{N}_{\text{UH}} \subseteq [N]$ such that $ |\mathcal{N}_{\text{HM}} |\ge N_r$,
the master chooses $N_r$ different $Y^{\boldsymbol{\nu}}_n $ from received messages  $ Y_{\mathcal{N}_{\text{HM}} }^{\boldsymbol{\nu}}$.  
Since any $N_r$-row submatrix of \(\mathbf{V}\) retains the structure of a Vandermonde matrix,    
the master can recover $ \left(\sum_{k=1}^K  W_{k,i} \right)_{i\in[N_r-T]}$ by using $N_r$  different $Y^{\boldsymbol{\nu}}_n $. 
By combining these results, the master can recover $ W = \sum_{k=1}^K W_{k}$.

\subsubsection*{Security Against Helpers}  
To provide a proof of security, we begin by presenting some useful properties of the messages defined above in the following lemmas. 

The following lemma presents some independence properties of the helper-side randomnesses, with its proof deferred to Subsection \ref{sec_Lamma1}. 
\begin{lemma} \label{lemma1} 
	For any helper set $\mathcal{S}_n \subseteq [N] \setminus \{n\}$,  the random variables defined in \eqref{ach_Zn} satisfy 
	\begin{subequations}
	\begin{align}   \label{lemma1_1}
	 	&  H \left(  \left(  Z_{\mathcal{S}_n ,n}^{(k)} \right)_{ n\in[N],k\in[K]} \right)   
	 	=  \sum_{n\in[N],k\in[K]} H  \left( Z_{\mathcal{S}_n,n}^{(k)} \right).  
	 \end{align}  
	Furthermore, if \( |\mathcal{S}_n| \le N_r - 1 \),  \(  Z^{(k)}_{\mathcal{S}_n,n}  \) contains $|\mathcal{S}_n|$ i.i.d. uniform variables, i.e.,     
	\begin{align}  \label{lemma1_2}
	H \left(  Z^{(k)}_{\mathcal{S}_n, n}    \right) = |\mathcal{S}_n|l =  \frac{L|\mathcal{S}_n|}{N_r-T}. 
	\end{align} 
	\end{subequations}  
\end{lemma} 

The following lemma shows that communication between helpers won't leak any user-to-helper messages other than those that the helpers are supposed to receive. 
The proof of Lemma \ref{lemma2} relies on the result of Lemma \ref{lemma1}, and is deferred to Subsection  \ref{sec_Lamma2}.  
\begin{lemma} \label{lemma2}  
		For    any $  \mathcal{T} \subseteq [N]$ where $|   \mathcal{T} | \le T$, and  any  $ \boldsymbol{\nu} \in 	\boldsymbol{\mathcal{N}}(N_r) $, we have  
		\begin{align*}    
		I \left (  X_{[K],[N] };     \mathcal{M}_{\mathcal{T}}^{ \boldsymbol{\nu}}   \big| X_{[K],\mathcal{T} }, Z_{\mathcal{T}}   \right) = 0. 
		\end{align*}   
\end{lemma}
 
Armed with Lemmas \ref{lemma1} and \ref{lemma2}, we now present the proof of security against helpers. 
First, consider the security after the users uploading phase. For any  $\mathcal{U} \subseteq [K]$,  and any $  \mathcal{T} \subseteq [N]$ where $|   \mathcal{T} | \le T$,  we have    
\begin{subequations}   
	\begin{align}  
		& I \left (  W_{[K]};  X_{[K],\mathcal{T} },    Z_{\mathcal{T}}  \Big| W_{\mathcal{U}}, F_{\mathcal{U}} \right)   \nonumber \\
		\label{seck1}
		\le & I \left (  W_{[K]};  X_{[K],\mathcal{T} },    \left( Q_{[N],[N_r-1]}^{(k)} \right)_{k \in [K]}  \bigg| W_{\mathcal{U}}, F_{\mathcal{U}} \right)    \\ 
		\label{seck2} 
		=  & I \left (  W_{[K]};  X_{[K],\mathcal{T} }       \big| W_{\mathcal{U}}, F_{\mathcal{U}} \right)    \\
		= 	  & H \left (  X_{[K],\mathcal{T} }       \big| W_{\mathcal{U}}, F_{\mathcal{U}} \right)  -  H \left (  X_{[K],\mathcal{T} }       \big| W_{[K]},  F_{\mathcal{U}} \right)  \nonumber   \\
		\label{seck3} 
		=    & H \left (  X_{[K] \setminus \mathcal{U} ,\mathcal{T} }       \big| W_{\mathcal{U}}, F_{\mathcal{U}} \right) \nonumber \\
		&  -  H \left( \left( \sum_{i=N_r - T +1}^{N_r}\alpha_{n}^{i-1}F_{k,i-N_r+T}^{\mathsf{T}} \right)   
		_{n\in \mathcal{T}, k \in [K] \setminus \mathcal{U}}       \right)    \\ 
		\label{seck4} 
		  \le & \frac{(K- |\mathcal{U}|)|\mathcal{T} |L}{N_r -T }-\frac{(K- |\mathcal{U}|)|\mathcal{T} |L}{N_r -T} = 0 ,  
	\end{align}       
\end{subequations}   
where \eqref{seck1}  follows from the fact that $Z_{n,[N]\setminus \{n\}}^{(k)}$, $n \in \mathcal{T}$, are functions of $ Q_{[N],[N_r-1]}^{(k)} $,  
\eqref{seck2} follows from the fact that $Q_{[N],[N_r-1]}^{(k)}$ are independent of $\left( W_{[K]}, F_{\mathcal{U}},X_{[K],\mathcal{T}} \right)$, 
\eqref{seck3} follows from the design of $X_{k,n}$, i.e., \eqref{X_kn},  and  the fact that $F_k$, $k \in [K]$, are mutually independent,  
the first term of \eqref{seck4} follows from the fact that each \(X_{k,n}\) has \(\frac{L}{N_r - T}\) symbols, and the uniform distribution maximizes entropy, 
and the second term of \eqref{seck4} follows from the fact that $  \sum_{i=N_r - T +1}^{N_r}\alpha_{n}^{i-1}F_{k,i-N_r+T}^{\mathsf{T}} 
=  \alpha_{n}^{N_r-T} \sum_{i=1}^{T} \alpha_{n}^{i-1}F_{k,i}^{\mathsf{T}} $, $n \in \mathcal{T}$,  are $|\mathcal{T}|$ linearly independent 
linear combinations of \(  F_{k,i}^{\mathsf{T}}, i \in [T] \), and that \(F_k\), \(k \in [K]\), are i.i.d. uniform random variables.

Finally, for any $\mathcal{U} \subseteq [K]$ ,  any $  \mathcal{T} \subseteq [N]$ where $|   \mathcal{T} | \le T$, and  any  $ \boldsymbol{\nu} \in 	\boldsymbol{\mathcal{N}}(N_r) $, we have  
\begin{subequations}  
\begin{align}  
	& I \left (  W_{[K]};  X_{[K],\mathcal{T} },    Z_{\mathcal{T}},\mathcal{M}_{\mathcal{T}}^{ \boldsymbol{\nu}}   \Big| W_{\mathcal{U}}, F_{\mathcal{U}} \right)  \nonumber \\ 
 = 	& I \left (  W_{[K]};  X_{[K],\mathcal{T} },    Z_{\mathcal{T}}  \Big| W_{\mathcal{U}}, F_{\mathcal{U}} \right)  \nonumber \\
 & + I \left (  W_{[K]}; \mathcal{M}_{\mathcal{T}}^{ \boldsymbol{\nu}}   \Big|  X_{[K],\mathcal{T} },    Z_{\mathcal{T}}, W_{\mathcal{U}}, F_{\mathcal{U}} \right) \nonumber \\
 \le 	&  0 + 
    I \left (  X_{[K],[N]}; \mathcal{M}_{\mathcal{T}}^{ \boldsymbol{\nu}}   \Big|  X_{[K],\mathcal{T} },    Z_{\mathcal{T}}, W_{\mathcal{U}}, F_{\mathcal{U}} \right)  \label{pfse1_a}\\
 \le 	&   
I \left (  X_{[K],[N]}; \mathcal{M}_{\mathcal{T}}^{ \boldsymbol{\nu}}   \Big|  X_{[K],\mathcal{T} },    Z_{\mathcal{T}} \right)  \label{pfse1_b} \\ 
 =  &  0,   \label{pfse1_c}
\end{align}      
\end{subequations}   
where  the first term of \eqref{pfse1_a} follows from \eqref{seck4},  
the second term of \eqref{pfse1_a} follows from the fact that $W_{[K]}$  can be decoded from $X_{[K],[N]}$ (see \eqref{X_kn}),     
\eqref{pfse1_b} follows from the fact that $W_{\mathcal{U}}$ and $F_{\mathcal{U}}$  can be decoded from $X_{[K],[N]}$ (see \eqref{X_kn}),   
and    \eqref{pfse1_c} follows from Lemma \ref{lemma2}. 
Since mutual information is greater than or equal to zero, from   \eqref{pfse1_c},  the proof of \eqref{security-a} is thus complete.

\subsubsection*{Security Against the Master}  We now prove that the above scheme guarantees security against the master, i.e., \eqref{security-b}. 
Consider $\mathcal{T}_0$ where $\mathcal{T} \subseteq \mathcal{T}_0$ and $|\mathcal{T}_0| = T$. Thus, for any $  \boldsymbol{\nu} \in \boldsymbol{\mathcal{N}}(N_r)$, we have  
 \begin{subequations}   
\begin{align}  
& H \left(   Y^{\boldsymbol{\nu}}_{[N]} \Big|W, X_{[K], \mathcal{T}_0} \right)  \nonumber \\ 
= & H \left(  \left( \sum_{k=1}^{K}  X_{k,n}  \right)_{n \in [N]}  \Bigg|\sum_{k=1}^{K}  W_{k} , X_{[K], \mathcal{T}_0} \right)    \label{pf2_YNa} \\
\le & H \left(    \left( \sum_{k=1}^{K}	
\sum_{i=N_r-T +1}^{N_r} \alpha_n^{i-1} F_{k,i+T-N_r}^{\mathsf{T}}  \right )_{n \in [N]} \right.  \nonumber \\  
&   \left. \quad \quad \left|  \sum_{k=1}^{K}  W_{k} , \left(  	
\sum_{k=1}^{K}   \sum_{i=N_r-T +1}^{N_r} \alpha_n^{i-1} F_{k,i+T-N_r}^{\mathsf{T}}  \right )_{n \in \mathcal{T}_0} \right. \right)    \label{pf2_YNb}   \\    
\le & H \left(  \left(
\sum_{i=N_r-T +1}^{N_r} \left(\alpha_n^{i-1}   \sum_{k=1}^{K}F_{k,i+T-N_r}^{\mathsf{T}} \right)  \right )_{n \in [N]} \right.  \nonumber \\  
&   \left. \quad \quad\quad \quad  \left|  \left(  	
      \sum_{k=1}^{K}  F_{k,i+T-N_r}^{\mathsf{T}}  \right)_{i  \in [N_r-T+1:N_r]} \right. \right)       \label{pf2_YNc} \\     
= &  0, \label{pf2_YNd}
\end{align}    
\end{subequations}    
where \eqref{pf2_YNa} follows from \eqref{correctness2}, 
\eqref{pf2_YNb} follows from the design of $X_{k,n}$, i.e., \eqref{X_kn},    
and \eqref{pf2_YNc} follows from the fact that $ \left( \sum_{i=N_r-T +1}^{N_r} \left(\alpha_n^{i-1}   \sum_{k=1}^{K}F_{k,i+T-N_r}^{\mathsf{T}} \right)  \right )_{n \in \mathcal{T}_0}$ contains $T$ linearly independent linear combinations of \( \sum_{k=1}^{K} F_{k,i}^{\mathsf{T}}, i \in [T] \), from which \( \sum_{k=1}^{K} F_{k,i}^{\mathsf{T}}, i \in [T] \) can be decoded. 

Armed with  \eqref{pf2_YNd}, the proof of \eqref{security-b} can be easily derived from the result of  \eqref{security-a}. 
More specifically, for any $\mathcal{U} \subseteq [K]$,  any $ \mathcal{T} \subseteq [N]$ where $|   \mathcal{T} | \le T$, and any $ \boldsymbol{\nu} \in \boldsymbol{\mathcal{N}}(N_r)$, we have 
\begin{subequations}  
\begin{align} 
	& I \left( W_{[K]}; Y^{\boldsymbol{\nu}}_{\mathcal{N}_{\text{UH}}},  X_{[K],\mathcal{T} },  Z_{\mathcal{T}},\mathcal{M}^{\boldsymbol{\nu}}_{\mathcal{T}}  \big | W, W_{\mathcal{U}}, F_{\mathcal{U}} \right)    \nonumber \\
	\le  & I \left(  W_{[K]}; Y^{\boldsymbol{\nu}}_{[N]},  X_{[K],\mathcal{T}_0 },  Z_{\mathcal{T}_0},\mathcal{M}^{\boldsymbol{\nu}}_{\mathcal{T}_0}  \Big | W, W_{\mathcal{U}}, F_{\mathcal{U}} \right)  \label{pf2_a}  \\  
	=  & I \left(  W_{[K]};  X_{[K],\mathcal{T}_0 },  Z_{\mathcal{T}_0},\mathcal{M}^{\boldsymbol{\nu}}_{\mathcal{T}_0}  \big | W, W_{\mathcal{U}}, F_{\mathcal{U}} \right) \label{pf2_b} \\ 
	\le  & I \left( W_{[K]};  X_{[K],\mathcal{T}_0 },  Z_{\mathcal{T}_0},\mathcal{M}^{\boldsymbol{\nu}}_{\mathcal{T}_0}  \big | W_{\mathcal{U}}, F_{\mathcal{U}}  \right )  \label{pf2_c} \\
	\le &  0,  \label{pf2_d}   
\end{align} 
\end{subequations}  
where \eqref{pf2_a} follows from $\mathcal{N}_{\text{UH}} \subseteq [N] $ and  $\mathcal{T} \subseteq \mathcal{T}_0$,  
\eqref{pf2_b} follows from \eqref{pf2_YNd}, 
\eqref{pf2_c} follows from  $W$ is a function of $W_{[K]}$, 
and \eqref{pf2_d} follows from the validity of \eqref{security-a}  in the case where   $ \mathcal{T} = \mathcal{T}_0$.    
Since mutual information is greater than or equal to zero, from   \eqref{pf2_d},  the proof of \eqref{security-b} is thus complete.  

\subsubsection*{Rate}
From the design of $X_{k,n}$ and $Y_n^{\boldsymbol{\nu}}$, i.e., \eqref{X_kn} and \eqref{Y_n}, we note that  $L_X = L_Y =l = \frac{L}{N_r-T}$.    
As a result, we have  
$(R_X,R_Y) = \left(\frac{1}{N_r-T},\frac{1}{N_r-T}\right) $ 
\begin{remark}
	In the proof above, we note that the independence and uniformity of the vectors \(W_1, W_2, \dots, W_K\) are not utilized, neither in the proof of correctness nor in the two security proofs. Thus, the rate tuple $(R_X,R_Y) = \left(\frac{1}{N_r-T},\frac{1}{N_r-T}\right) $ is achievable for \(W_1, W_2, \dots, W_K\) with arbitrary distributions.
\end{remark}

Hence, the proof of achievability is complete, provided that we prove Lemmas \ref{lemma1} and \ref{lemma2}, which will be established next.

\subsection{Proof of Lemma 1}\label{sec_Lamma1}
	First, consider \eqref{lemma1_1}.   
	Since for any $k\in[K]$ and any $ n\in[N]$,
	$\left(Z_{i, n}^{(k)}\right)_{ i \in [N] \setminus\{n\}} $ is a function of $Q_{n,[N_r-1]}^{(k)}$, 
	and  $Q_{n,[N_r-1]}^{(k)}$, $ k \in [K]$, $n \in [N]$, are i.i.d. uniform variables,  $Z_{[N] \setminus \{n\} ,n}^{(k)}$, $ k \in [K]$, $n \in [N]$, are mutually independent. 
	Furthermore, since \(\mathcal{S}_n \subseteq [N] \setminus \{n\}\), the proof of \eqref{lemma1_1} is thus completed.
	
	Next, consider  \eqref{lemma1_2}. For any   $\mathcal{S}_n \subseteq [N] \setminus \{n\}$ such that  \( |\mathcal{S}_n| \le N_r - 1 \),  we have \begin{subequations}    
 	\begin{align} 
 	& \left[ \left(Z_{i,n}^{(k)} \right)_{i \in \mathcal{S}_n }    \right]^{\mathsf{T}}     \nonumber\\ 
 =   &  \left[ \left( \mathbf{e}_i^{N} \right)_{ i \in \mathcal{S}_n   }   \right] ^{\mathsf{T}}     \mathbf{S}_n    \overset{\sim}{\mathbf{G}}   
 	\left[   
 	Q_{n,1}^{(k)}, Q_{n,2}^{(k)},   \cdots,  Q_{n,N_r-1}^{(k)}  
 	\right]^{\mathsf{T}}    \label{lemma1_21a}   \\    
 	=   & \left[ \left( \mathbf{e}_i^{N} \right)_{ i \in \mathcal{S}_n   }   \right] ^{\mathsf{T}}     \mathbf{S}_n      \left[ \left(\mathbf{e}_i^{N_r}\right)_{ i \in [2:N_r]} \right] \left[ \left(\mathbf{e}_i^{N_r}\right)_{ i \in [2:N_r]} \right]^{\mathsf{T}}  \overset{\sim}{\mathbf{G}}       \nonumber  \\    &     \quad   \quad \quad \quad \quad \quad \quad \quad 
  \left[   
 Q_{n,1}^{(k)}, Q_{n,2}^{(k)},   \cdots,  Q_{n,N_r-1}^{(k)}  
 \right]^{\mathsf{T}}   ,   \label{lemma1_21b}  
 \end{align} 
\end{subequations}    
where \eqref{lemma1_21a}  follows from the definition of $Z_{i,n}^{(k)}$, i.e.,  \eqref{ach_Zn},    
\eqref{lemma1_21b} follows from the fact that the first line of $\overset{\sim}{\mathbf{G}}$ is all zero.

By setting $\mathcal{N} =  \mathcal{S}_n \cup \{n\}$ in \eqref{rank0}, we have
\begin{align*} 
	& \text{rank} \left(  \left[ \left( \mathbf{e}_i^{N} \right)_{ i \in   \mathcal{S}_n \cup \{n\}   }   \right] ^{\mathsf{T}}     \mathbf{S}_n    \right)     =  | \mathcal{S}_n | + 1,      
\end{align*}  
which implies that $\left[ \left( \mathbf{e}_i^{N} \right)_{ i \in   \mathcal{S}_n \cup \{n\}   }   \right] ^{\mathsf{T}}     \mathbf{S}_n   $ is  full rank. 
From Remark \ref{remarkZ}, we have  
$	\left( 	\mathbf{e}_n^{N}  \right)^{\mathsf{T}} \mathbf{S}_n  =   (1,\mathbf{0}_{1\times (N_r-1)})  $. Thus,   $ \left[ \left( \mathbf{e}_i^{N} \right)_{ i \in \mathcal{S}_n   }   \right] ^{\mathsf{T}}     \mathbf{S}_n      \left[ \left(\mathbf{e}_i^{N_r}\right)_{ i \in [2:N_r]} \right]$ can be seen as a submatrix  of  $\mathbf{S}_n$  and has  full rank, i.e., 
\begin{align} 
		\text{rank} \bigg( \left[ \left( \mathbf{e}_i^{N} \right)_{ i \in \mathcal{S}_n   }   \right] ^{\mathsf{T}}     \mathbf{S}_n      \left[ \left(\mathbf{e}_i^{N_r}\right)_{ i \in [2:N_r]} \right]   \bigg)  =  | \mathcal{S}_n   |. \label{lemma1_22} 
\end{align} 
	By combining the results of \eqref{lemma1_21b} and \eqref{lemma1_22},  we note that $Z_{\mathcal{S}_n,n}^{(k)} $ contains $|\mathcal{S}_n|$  linearly independent linear combinations of   $Q_{n,[N_r-1]}^{(k)}$. 
	Since $Q_{n,[N_r-1]}^{(k)}$ contains   $N_r-1$ i.i.d. uniform $l  \times 1$ vectors over $\mathbb{F}_q^{l \times 1}$, we have   
 \begin{align*}  
 	H \left(  Z^{(k)}_{\mathcal{S}_n, n}    \right) =  |\mathcal{S}_n| l = \frac{ |\mathcal{S}_n|L}{N_r-T}. 
 \end{align*}   
 The proof of Lemma \ref{lemma1} is thus complete.  

\subsection{Proof of Lemma 2}\label{sec_Lamma2}  
For any $k \in [K]$,  any $ \boldsymbol{\nu} \in 	\boldsymbol{\mathcal{N}}(N_r) $, and any $ \mathcal{T} \subseteq [N]$, $|\mathcal{T}| \le T$, let $ \mathcal{T}_k = \mathcal{T} \setminus \mathcal{N}_k$.   Then, for any $t \in \mathcal{T}_k$, considering the messages between helpers used to recover $X_{k,t}$, we have 
\begin{subequations}  
\begin{align} 
	& H \left(   M_{\mathcal{N}_k,t}^{(k)}     \Big| X_{[K],\mathcal{T} }, Z_{\mathcal{T}}     \right)  \nonumber \\  
	 \label{pfl2a}
	 \le & H \left (    \left(X_{k,n} + Z_{n,t}^{(k)}\right)_{ n \in \mathcal{N}_k }     \bigg|  \left(X_{k,n} + Z_{n,t}^{(k)} \right)_{ n \in \mathcal{T}  }     \right)  \\   
     = & H \left(     \left(X_{k,n} + Z_{n,t}^{(k)}\right)_{ n \in \mathcal{N}_k \cup   \mathcal{T}}        \right)  \nonumber  \\
     &   - H \left(  X_{k,t},     \left(X_{k,n} + Z_{n,t}^{(k)} \right)_{ n \in \mathcal{T} \setminus \{t\} }     \right) \label{pfl2b}  \\ 
     \le & H \left(  X_{k,t}\right) +  (N_r-1)l
          - H \left(  X_{k,t} \right) \nonumber \\
     &  -   H \left(   \left(X_{k,n} + Z_{n,t}^{(k)} \right)_{ n \in \mathcal{T} \setminus \{t\} }   \bigg|  X_{k,[N]}   \right)  \label{pfl2c} \\ 
     =  &    (N_r-1)l   - (|\mathcal{T}|-1)l  = (N_r-|\mathcal{T}|)l, \label{pfl2d} 
\end{align}  
\end{subequations}  
where \eqref{pfl2a} follows from  the design of $M_{n,i}^{(k)}$, i.e., \eqref{ach_M0},  and the fact that  $ \left(X_{k,n} + Z_{n,t}^{(k)} \right)_{ n \in \mathcal{T} }$ can be  determined by  $X_{[K],\mathcal{T}}$ and $ Z_{\mathcal{T}}$ (see   \eqref{ach_Zn} and Remark \ref{remarkZ}),    
\eqref{pfl2b}  follows from $Z_{t,t}^{(k)} =  \mathbf{0}_{l \times 1 }$,  
\eqref{pfl2c} follows from the fact that \(\left(X_{k,n} + Z_{n,t}^{(k)}\right)_{n \in [N]}\) contains at most \(N_r\) different linear combinations of \(W_{k,[N_r-T]}\), \(F_{k,[T]}\), and \(Q^{(k)}_{t,[N_r-1]}\), one of which is \(X_{k,t}\), and the message length of each linear combination is \(l\) (see \eqref{design}),  
and \eqref{pfl2d}  follows from Lemma \ref{lemma1}, more specifically, $ H \left( Z_{\mathcal{T} \setminus \{t\}, t}^{(k)}\right)  =  (|\mathcal{T}| -1  ) l$. 
 
Finally, for  any $ \boldsymbol{\nu} \in 	\boldsymbol{\mathcal{N}}(N_r) $, and any $ \mathcal{T} \subseteq [N]$, $|\mathcal{T}| \le T$, we have   
\begin{subequations}  
\begin{align}  
	& I \left (  X_{[K],[N] };     \mathcal{M}_{\mathcal{T}}^{ \boldsymbol{\nu}}   \big| X_{[K],\mathcal{T} }, Z_{\mathcal{T}}   \right)  \nonumber \\
	=	& H \left (    \mathcal{M}_{\mathcal{T}}^{ \boldsymbol{\nu}}   \big| X_{[K],\mathcal{T} }, Z_{\mathcal{T}}   \right)  - H \left (     \mathcal{M}_{\mathcal{T}}^{ \boldsymbol{\nu}}   \big| X_{[K],[N]}, Z_{\mathcal{T}}   \right)  \nonumber \\
	\le	& \sum_{k \in [K]} \sum_{t \in  \mathcal{T}_k} H \left (   M_{\mathcal{N}_k,t}^{(k)}     \Big| X_{[K],\mathcal{T} }, Z_{\mathcal{T}}    \right)  \nonumber \\ 
	& - H \left (    \left(Z_{\mathcal{N}_k,\mathcal{T}_k}^{(k)} \right)_{  k \in [K] } \bigg| \left( Z_{n,[N]\setminus{\{n\}}}^{(k)}  \right)_{ n \in \mathcal{T}, k \in [K] }    \right) \label{pfl3a}    \\ 
	\le	& \sum_{k \in [K]} | \mathcal{T}_k| (N_r-|\mathcal{T}|)l      - \sum_{k \in [K]}  \sum_{t \in \mathcal{T}_k}  H \left(Z_{\mathcal{N}_k,t}^{(k)}       \Big|  Z_{ \mathcal{T} \setminus \{t\},t}^{(k)}   \right) \label{pfl3b}   \\  
	= & \sum_{k \in [K]}| \mathcal{T}_k| (N_r-|\mathcal{T}|)l    \nonumber \\
	& - \sum_{k \in [K]}  \sum_{t \in \mathcal{T}_k}   \left( H \left(Z_{\mathcal{N}_k \cup \mathcal{T} \setminus \{t\},t}^{(k)}      \right)    - H \left(  Z_{ \mathcal{T} \setminus \{t\},t}^{(k)} \right)    \right)    \nonumber  \\     
	= & \sum_{k \in [K]}  |\mathcal{T}_k| (N_r-|\mathcal{T}|)l     - \sum_{k \in [K]}  |\mathcal{T}_k| (N_r-|\mathcal{T}|)l  \label{pfl3c}  \\
	= & 0,    \label{pfl3d} 
\end{align}  
\end{subequations}        
where \eqref{pfl3a} follows from  the design of $  \mathcal{M}_{\mathcal{T}}^{ \boldsymbol{\nu}}$ and $Z_{\mathcal{T}}$, i.e., \eqref{ach_M} and \eqref{ach_Zn},  and the fact that $\left(Z_{\mathcal{N}_k,\mathcal{T}_k}^{(k)} \right)_{k \in [K]}$ is independent of $X_{[K],[N]}$,   
the first term of \eqref{pfl3b} follows from  \eqref{pfl2d}, 
the second term of \eqref{pfl3b} follows from   \eqref{lemma1_1}, 
and \eqref{pfl3c} follows from \eqref{lemma1_2}.  

Since mutual information is greater than or equal to zero, from   \eqref{pfl3d}, we have
\begin{align*}  
& I \left (  X_{[K],[N] };     \mathcal{M}_{\mathcal{T}}^{ \boldsymbol{\nu}}   \big| X_{[K],\mathcal{T} }, Z_{\mathcal{T}}   \right)  =  0,     
\end{align*}  
The proof of Lemma \ref{lemma2} is thus complete.

\section{Converse Proof} \label{sec5}
In this section, we provide the converse proof of Theorem \ref{Theo1}. 
The proof is divided into two cases: $N_r \le T$ and $N_r > T$. For the case $N_r \le T$, we prove that no achievable scheme exists. For the case $N_r > T$, we derive the lower bounds for the communication rates from the user to the helper and from the helper to the master, respectively.

\subsection{$ N_r \le T$: Proof of $\mathcal{R}^* = \emptyset $}\label{subsection1}
When \( N_r \le T \), a contradiction arises between the security constraints and the correctness constraint. 
To derive this contradiction, we consider the case \( \mathcal{T} = \mathcal{N}_{\text{HM}} = [N_r] \)  and  $\mathcal{U} = \emptyset$. Since \( N_r \le T \le N \), the choice of \( \mathcal{T} \) and \( \mathcal{N}_{\text{HM}} \) is feasible.  
When    \( \mathcal{T} = \mathcal{N}_{\text{HM}} = [N_r] \) and   $\mathcal{U} = \emptyset$,  from the security against helpers, i.e., \eqref{security-a}, we have 
\begin{subequations}  
\begin{align}
0 = & I \left(  W_{[K]} ; X_{[K],[N_r]},Z_{[N_r]},\mathcal{M} _{[N_r]}^{\boldsymbol{\nu} }  \right) \nonumber \\
\ge   & I \left( W; Y_{[N_r]}^{\boldsymbol{\nu} }  \right) \label{con1_1} \\
= & H(W) - H \left(W \Big|Y_{[N_r]}^{\boldsymbol{\nu} } \right) \nonumber \\
= & L-0 = L, \label{con1_2} 
\end{align} 
\end{subequations}
where \eqref{con1_1} follows from the fact that \( Y_{[N_r]}^{\boldsymbol{\nu} } \) can be determined by \( X_{[K],[N_r]} \), \( \mathcal{M}_{[N_r]}^{\boldsymbol{\nu} } \), and \( Z_{[N_r]} \) (see \eqref{Yn}), and that \( W \) is a function of \( W_{[K]} \),   
the first term of \eqref{con1_2} follows from the fact that \( W_{[K]} \) consists of \( K \) i.i.d. random vectors, each uniformly distributed on \( \mathbb{F}_q^{L \times 1} \), and that \( W \) is a linear combination of \( W_{[K]} \),   
and the second   term of \eqref{con1_2} follows from the correctness constraint, i.e., \eqref{correctness}.  

Thus, we arrive at a contradiction, i.e., $ 0 \ge  L$.  
As a result, there is no achievable scheme, i.e., \( \mathcal{R}^* = \emptyset \).   
The proof of $ \mathcal{R}^* = \emptyset $ is thus complete. 

\subsection{$ N_r > T$: Proof of $R_X \ge \frac{1}{N_r-T} $} \label{subsection2}
In this subsection, for the case $N_r > T $, we derive the lower bound for the rate of the user-to-helper communication. 

We start with the following lemma, which shows that the local gradient of user $k$ can be recovered from all the messages successfully uploaded by user $k$.  
\begin{lemma} 
	\label{lem-3}
	For any $k \in[K]$ and $\mathcal{N}_k \subseteq [N] $ such that $|\mathcal{N}_k| \ge N_r$, the following  equality holds:  
	\begin{align*}
	I  \left( W_k ; X_{k,\mathcal{N}_k} \right) = L. 
	\end{align*} 
\end{lemma}
\begin{IEEEproof}   
	From the correctness constraint, i.e., \eqref{correctness}, 
	we have 
\begin{subequations}	
	\begin{align}   
		0 = & H \left(W \big|Y^{\boldsymbol{\nu} }_{\mathcal{N}_{\text{HM}}} \right) 
		\ge  H \left(W  \big|Y^{\boldsymbol{\nu} }_{\mathcal{N}_{\text{HM}}}, W_{[K] \setminus \{k\}} \right)   \nonumber \\
		\label{lem3_1}  
		=  &   H \left( W_k \big| Y^{ \boldsymbol{\nu}}_{\mathcal{N}_{\text{HM}}} , W_{[K] \setminus \{k\}} \right)   \\   
		\ge   &  H \left( W_k \Big|  \left(X_{\mathcal{K}_n, n }\right)_{n \in \mathcal{N}_{\text{HM}} }  , Z_{\mathcal{N}_{\text{HM}}},\mathcal{M}^{\boldsymbol{\nu}}_{\mathcal{N}_{\text{HM}}},  W_{[K] \setminus \{k\}} \right)   \label{lem3_2}   
		\\   
		\ge    &  H \left( W_k \Big|\left(X_{\mathcal{K}_n, n }\right)_{n \in [N] } ,Z_{[N]},  W_{[K] \setminus \{k\}} \right)   \label{lem3_3} 
		 \\  
		\ge &  H \left( W_k \big|X_{k,\mathcal{N}_k},Z_{[N]},  W_{[K] \setminus \{k\}},  F_{[K] \setminus \{k\}} \right)    \label{lem3_4} \\
		= &  H \left( W_k \big|X_{k,\mathcal{N}_k} \right) \label{lem3_5}, 
	\end{align} 
\end{subequations}
	where \eqref{lem3_1} follows from  $ W = \sum_{k = 1}^{K} W_k$, 
	\eqref{lem3_2} follows from the fact that 
	\( Y^{ \boldsymbol{\nu}  }_{\mathcal{N}_{\text{HM}}} \) can be determined by 
	 $ \left(X_{\mathcal{K}_n, n }\right)_{n \in \mathcal{N}_{\text{HM}} } $,  $Z_{\mathcal{N}_{\text{HM}}}$ and  $\mathcal{M}^{\boldsymbol{\nu}   }_{\mathcal{N}_{\text{HM}}} $  (see \eqref{Yn}),    
	\eqref{lem3_3} follows from $\mathcal{N}_{\text{HM}} \subseteq [N]$, and the fact that  $\mathcal{M}^{\boldsymbol{\nu}}_{\mathcal{N}_{\text{HM}}}$ 
	can be determined by $ \left(X_{\mathcal{K}_n, n }\right)_{n \in [N] }$ and $Z_{[N]}$ (see \eqref{defMn}),  
	\eqref{lem3_4} follows from  the fact that  $\left(X_{i,\mathcal{N}_i}\right)_{i \in [K] \setminus \{k\}}$
	can be determined by $W_{[K] \setminus \{k\}}$ and $F_{[K] \setminus \{k\}}$ (see \eqref{defXkn}),   
	and \eqref{lem3_5} follows from the fact that $X_{k, \mathcal{N}_k} $ is a function of $W_k$ and $F_k$, and that
	$\left( W_{k},F_k\right)$  is independent of  $\left( Z_{[N]}, W_{[K]\setminus \{k\}},F_{[K]\setminus \{k\}}\right)$.  
	Thus, we have 
	\begin{align*}
		& I  \left( W_k ; X_{k,\mathcal{N}_k} \right)  
		=   H  \left( W_k \right) -   H  \left( W_k | X_{k,\mathcal{N}_k} \right) \nonumber =   L,  
	\end{align*}  
	where the last step follows from the assumption that $W_k$ is a $L \times 1$   uniform  vector, i.e., \eqref{Wk}.  
\end{IEEEproof}

The above lemma is a consequence of the correctness constraint. Furthermore, consider the security against helpers, then for any $k \in[K]$, we have 
\begin{align} \label{proofR10}
 & I \left( W_k;X _{k,[T]}\right)  \nonumber   \\
\le  &  I \left(W_{[K]}; X_{[K], [T]},\mathcal{M}_{[T]}^{\boldsymbol{\nu} } ,
 Z_{[T]}  \right)  =  0, 
\end{align}  
where \eqref{proofR10} follows from \eqref{security-a} with  $\mathcal{T} =[T]$ and $\mathcal{U} = \emptyset$.

Armed with  Lemma \ref{lem-3} and \eqref{proofR10}, consider $ \mathcal{N}_k = [N_r]$, then we have 
\begin{subequations} 
\begin{align}  
L = &  I  \left( W_k; X_{k,[N_r]} \right) \nonumber \\
= & I \left( W_k; X_{k,[T]}\right) + I \left( W_k;X_{k,[N_r] \setminus [T]} \big | X_{k,[T]}\right)   \nonumber  \\ 
= & I \left( W_k;X_{k,[N_r] \setminus [T]}  \big| X_{k,[T]}\right)   \label{proofR1:1} \\ 
\le &  H \left(   X_{k,[N_r] \setminus [T]}\right) 
\le \sum_{n\in [N_r] \setminus [T]} H\left(X_{k,n} \right)\nonumber \\
\le & (N_r-T)L_X,  \label{proofR1:2}
\end{align}   
\end{subequations} 
where \eqref{proofR1:1} follows from  \eqref{proofR10}, 
and \eqref{proofR1:2} follows from  the fact that each $X_{k,n}$ contains at most $L_X$ symbols. 


Thus, following from the definition of $R_X$, we have 
\begin{align*}
	 R_X = & \frac{L_X }{L} \ge \frac{1}{N_r-T}. 
\end{align*}
The proof of $ R_X \ge \frac{1}{N_r-T} $ when $ N_r > T$ is thus complete.

\subsection{$ N_r > T$: Proof of $R_Y \ge \frac{1}{N_r-T}$} \label{subsection3}
In this subsection, for the case $N_r > T $, we derive the lower bound for the rate of the helper-to-master communication. 
 
Consider $\mathcal{N}_{\text{UH}} = \mathcal{N}_{\text{HM}} = [N_r]$, then we have   
\begin{subequations}  
\begin{align} 
	 \label{proofR2_a} 
L = & H \left(W \right)  \\ 
 \label{proofR2_b}  %
= &  H \left(W\right) - H \left(W \Big| Y_{[N_r]}^{\boldsymbol{\nu} }\right)    
  =  I \left(W ; Y_{[N_r]}^{\boldsymbol{\nu} }\right) \\
= & I \left(W ; Y_{[T]}^{\boldsymbol{\nu} }\right)
+   I\left( W ;  Y_{[N_r]\setminus[T]}^{\boldsymbol{\nu} } 
\Big | Y_{[T]}^{\boldsymbol{\nu} }\right)  \nonumber \\
\le &  0 + H \left( Y_{[N_r]\setminus[T]}^{\boldsymbol{\nu} } \right) \label{proofR2_c} \\
\le & \sum_{n\in[N_r]\setminus[T]} H\left( Y_n^{\boldsymbol{\nu} } \right) 
\le   (N_r-T)L_Y,\label{proofR2_d}  
\end{align}
\end{subequations} 
where \eqref{proofR2_a} follows from $W = \sum_{k=1}^{K} W_k$ and the assumption that $W_{[K]}$ are $K$ i.i.d. $L \times 1$ uniform  vectors,   
\eqref{proofR2_b} follows from the correctness constraint, i.e., \eqref{correctness},   
 \eqref{proofR2_c} follows from $W = \sum_{k=1}^{K} W_k$, \eqref{Yn}, and the security against helpers, i.e., \eqref{security-a} with $\mathcal{T} = [T]$, $\mathcal{N}_{\text{UH}} = [N_r]$  and   $\mathcal{U} = \emptyset$,  
 and  \eqref{proofR2_d} follows from the fact that each $Y_n^{ \boldsymbol{\nu} }$ contains at most $L_Y$ symbols. 
 
Thus, following from the definition of $R_Y$, we have  
\begin{align*}
  R_Y  & = \frac{L_Y}{L} \ge \frac{1}{N_r-T}. 
\end{align*} 
The proof of $ R_Y \ge \frac{1}{N_r-T} $ when $ N_r > T$ is thus complete.

\section{Conclusions} \label{sec6}
We propose and study a hierarchical secure coded gradient aggregation problem, where users communicate with the master through an intermediate layer of helpers that can communicate with each other. 
For straggling communication links with a resiliency threshold of \( N_r \), the goal of the server is to recover the summation of all user gradients.  
With at most \( T \) colluding helpers and any number of colluding users, the security constraints require that no information about \( W_{[K]} \) is leaked beyond what is already known by the colluding users, or what is contained in the required summation.  
We propose an achievable scheme along with a matching converse bound, demonstrating that secure aggregation is infeasible when \( N_r \leq T \), whereas when \( N_r > T \), both the user-to-helper and helper-to-master optimal communication rates are \( \frac{1}{N_r - T} \).

\appendices  
\bibliographystyle{IEEEtran}
\bibliography{ref}

\end{document}